\documentclass[%
 reprint,
 superscriptaddress,
showpacs,preprintnumbers,
nofootinbib,
 amsmath,amssymb,
 aps,prc,
floatfix,
]{revtex4-1}

\usepackage{siunitx}
\usepackage{graphicx}
\usepackage{dcolumn}
\usepackage{bm}
\usepackage{todonotes}
\usepackage{wrapfig}
\usepackage{rotating}
\usepackage{epstopdf}

\usepackage[normalem]{ulem}
\newcommand{\soutthick}[1]{%
    \renewcommand{\ULthickness}{0.9pt}%
       \sout{#1}%
    \renewcommand{\ULthickness}{0.4pt}
}

\begin{document}

\title{$^{145}$Ba and $^{145, 146}$La structure from lifetime measurements}

\author{B.~Olaizola}
\email{bruno.olaizola@cern.ch}
\altaffiliation{Present address: ISOLDE-EP, CERN, CH-1211 Geneva 23, Switzerland}
\affiliation{TRIUMF, 4004 Wesbrook Mall, Vancouver, BC, V6T 2A3, Canada}

\author{A.~Babu}
\affiliation{TRIUMF, 4004 Wesbrook Mall, Vancouver, BC, V6T 2A3, Canada}

\author{R.~Umashankar}
\affiliation{TRIUMF, 4004 Wesbrook Mall, Vancouver, BC, V6T 2A3, Canada}
\affiliation{Department of Physics and Astronomy, University of British Columbia, Vancouver, British Columbia V6T 1Z4, Canada}

\author{A.B.~Garnsworthy}
 \affiliation{TRIUMF, 4004 Wesbrook Mall, Vancouver, BC, V6T 2A3, Canada}

\author{G.C.~Ball}
 \affiliation{TRIUMF, 4004 Wesbrook Mall, Vancouver, BC, V6T 2A3, Canada}

\author{V.~Bildstein}
\affiliation{Department of Physics, University of Guelph, Guelph, ON, N1G 2W1, Canada}

\author{M.~Bowry} 
\altaffiliation{Present address: School of Engineering, Computing and Physical Sciences, University of the West of Scotland, High Street, Paisley PA1 2BE, United Kingdom}
\affiliation{TRIUMF, 4004 Wesbrook Mall, Vancouver, BC, V6T 2A3, Canada}

\author{C.~Burbadge}
\affiliation{Department of Physics, University of Guelph, Guelph, ON, N1G 2W1, Canada}

\author{R.~Cabellero-Folch} 
\affiliation{TRIUMF, 4004 Wesbrook Mall, Vancouver, BC, V6T 2A3, Canada}

\author{I.~Dillmann} 
\affiliation{TRIUMF, 4004 Wesbrook Mall, Vancouver, BC, V6T 2A3, Canada}
\affiliation{Department of Physics and Astronomy, University of Victoria, Victoria, British Columbia V8P 5C2, Canada}

\author{A.~Diaz-Varela}
\affiliation{Department of Physics, University of Guelph, Guelph, ON, N1G 2W1, Canada}

\author{R.~Dunlop}
\affiliation{Department of Physics, University of Guelph, Guelph, ON, N1G 2W1, Canada}

\author{A.~Estrad{\'e}}
\affiliation{Department of Physics, Central Michigan University, Mount Pleasant, MI 48859, USA}

\author{P.E.~Garrett}
\affiliation{Department of Physics, University of Guelph, Guelph, ON, N1G 2W1, Canada}

\author{G.~Hackman}
\affiliation{TRIUMF, 4004 Wesbrook Mall, Vancouver, BC, V6T 2A3, Canada}

\author{A.D.~MacLean}
\affiliation{Department of Physics, University of Guelph, Guelph, ON, N1G 2W1, Canada}

\author{J.~Measures} 
\affiliation{TRIUMF, 4004 Wesbrook Mall, Vancouver, BC, V6T 2A3, Canada}
\affiliation{Department of Physics, University of Surrey, Guildford, Surrey, GU2 7XH, United Kingdom}

\author{C.J.~Pearson} 
\affiliation{TRIUMF, 4004 Wesbrook Mall, Vancouver, BC, V6T 2A3, Canada}

\author{B.~Shaw}
 \affiliation{TRIUMF, 4004 Wesbrook Mall, Vancouver, BC, V6T 2A3, Canada}

\author{D.~Southall}
\altaffiliation{Present Address: Department of Physics, University of Chicago, Chicago, Illinois 60637, USA}
 \affiliation{TRIUMF, 4004 Wesbrook Mall, Vancouver, BC, V6T 2A3, Canada}

\author{C.E.~Svensson}
\affiliation{Department of Physics, University of Guelph, Guelph, ON, N1G 2W1, Canada}

\author{J.~Turko}
\affiliation{Department of Physics, University of Guelph, Guelph, ON, N1G 2W1, Canada}

\author{K.~Whitmore} 
\affiliation{Department of Chemistry, Simon Fraser University, Burnaby, British Colombia V5A 1S6, Canada}

\author{T.~Zidar}
\affiliation{Department of Physics, University of Guelph, Guelph, ON, N1G 2W1, Canada}

\date{\today}

\begin{abstract}

The occurrence of octupole shapes in even-mass neutron-rich Ba isotopes has been well established. However, the situation with the odd-mass Ba and odd or odd-odd La nuclei around them is far from settled. In order to shed light on these less-studied isotopes, a fast-timing experiment was performed using GRIFFIN at TRIUMF-ISAC. A wealth of excited-state lifetimes in the 100~ps to few ns range have been measured in $^{144, 145, 146}$Ba and $^{145,146}$La populated in the $\beta^-$ and $\beta^--n$ decay of $^{145,146}$Cs. The results do not allow to draw firm conclusions on the possible octupole deformation of these nuclei but suggest different spin and parity assignments than previous works. This work highlights the need for more detailed study of the odd and odd-odd isotopes in this region to properly understand their structure.

\end{abstract}

\pacs{Valid PACS appear here}
\maketitle

\section{\label{sec:intro}Introduction}

The ground states for most nuclei across the chart of nuclides are dominated by prolate deformed shapes with pockets of sphericity observed near doubly-magic nuclei~\cite{Pritychenko2016,Campbell2016}. In addition, there has long been discussed a handful of regions where stable reflection-asymmetric octupole deformation or a dynamic octupole vibration may become the most favored geometric configuration~\cite{Butler1996}. These regions are located around the so-called octupole magic numbers, which are nucleon numbers where closed shells have single-particle orbitals with $\Delta n=1$, $\Delta l=3$, and $\Delta j=3$ coinciding at the Fermi surface ($n$ is the shell number, $l$ is the orbital angular momentum and $j$ is the spin-orbit one). Thus, $^{144}$Ba ($Z=56,N=88$) is considered a doubly magic octupole nucleus since the orbitals for both neutrons ($\nu d_{5/2}-h_{11/2}$) and protons ($\pi f_{7/2}-i_{13/2}$) satisfy this condition. Therefore, the region around it is expected to present strong octupole correlations.


A number of experiments utilizing~\cite{Hamilton1995,Zhu1995,Urban1997,Zhu1999Ba,Naidja2017, Lica2018,Zhu2020} identified ground-state bands in the even-even Ba isotopes with interleaved even-spin, positive-parity and odd-spin, negative-parity states connected by enhanced $E1$ transitions which provides indirect evidence for strong octupole correlations~\cite{Butler1996}. Only recently, Coulomb excitation of radioactive beams of even-even $^{144}$Ba~\cite{Bucher2016} and $^{146}$Ba~\cite{Bucher2017} have found direct evidence of octupole deformation through the measurement of large $B(E3; 3^-_1 \rightarrow 0^+_1)$ values. These fast $E3$ transitions observed, $B(E3)\sim50$~W.u. for both nuclei, are consistent with permanent octupole deformation in the ground state. 


Although intrinsic reflection asymmetry was theoretically predicted for $^{145}$Ba~\cite{Leander1985}, fission experiments seem to discard permanent octupole deformation of the ground state of this nucleus, due to the non-observation of interlacing opposite-parity bands. Nevertheless, these same experiments found indirect hints that octupole vibrations could occur at intermediate values of angular momentum~\cite{Zhu1999Ba, Rzaca2012}. This lack of ground-state octupole deformation was later expanded to $^{147}$Ba~\cite{Rzaca2013}.



Similar studies have been performed for the lanthanum $Z$=57 isotopes. Fission experiments found alternating-parity bands at intermediate and high spins for $^{143, 145, 147}$La, but the non-observation of parity doublets for the ground state prevented the establishment of permanent octupole deformation~\cite{Urban1996, Zhu1999La, Wang2007, Luo2009}. As is common in nuclear physics, odd-odd nuclei are typically less well studied than their even-even or odd-$A$ neighbours. Very limited information on the levels or band structure of $^{146}$La is available, with only indirect hints of octupole correlations extracted from the electric dipole moment $(D_0)$ values~\cite{Wang2017}.

In order to study low lying states in $^{145,146}$Ba, a $\beta$-decay experiment was performed using the GRIFFIN spectrometer at the TRIUMF-ISAC facility. A detailed study of the $\beta$ and $\beta n$ decays of $^{145,146}$Cs will be published elsewhere. In this article, we report new lifetime measurements for excited states in $A=145,146$ isotopes of Ba and their decay product, La. This provided information not just on the collective transitions which make up the rotational bands built on the ground-state configuration, but also an assessment of the single-particle strengths between different configurations in this region.

\section{\label{sec:expsetup}Experimental setup and methods}

Isotopes of cesium were produced from reactions induced in an uranium carbide target by a 9.7\,$\mu$A, 478\,MeV proton beam delivered by the TRIUMF Cyclotron~\cite{Bylinskii2013}. The Cs atoms created in the target that diffused out of the material were ionized and accelerated to 28\,keV, mass separated and delivered to the experimental station.

The ions were stopped in a mylar tape at the central focus of the Gamma-Ray Infrastructure For Fundamental Investigations of Nuclei (GRIFFIN) spectrometer~\cite{Svensson2014,Garnsworthy2019}. In this case, 12 GRIFFIN high-purity Germanium (HPGe) clover detectors~\cite{Rizwan2016} placed at a distance of 11\,cm from the implantation point were used for the detection of $\gamma$ rays. Ten plastic scintillator paddles of the SCintillating Electron-Positron Tagging ARray (SCEPTAR) array were located in the upstream part of the chamber for $\beta$-particle detection. The Zero-Degree Scintillator (ZDS), a single 1~mm thick fast plastic disc was positioned very close to the implantation point at zero-degrees to the beam axis also for the detection of $\beta$ particles. The DEuterated Scintillator Array for Neutron Tagging (DESCANT)~\cite{Garrett2014,Bildstein2013,Bildstein2015} was installed in the downstream location but the data from these detectors were not used in this analysis. Four cylindrical $5.1 \;\text{cm} \times 5.1 \;\text{cm}$ 5\% cerium-doped lanthanum bromide [LaBr$_3$(Ce)] scintillators were installed in the ancillary detector positions of the array at a distance of 12.5\,cm from the implantation point. No BGO shielding was available at the time of this experiment.

\begin{figure*}
\includegraphics[width=0.81\textwidth, keepaspectratio]{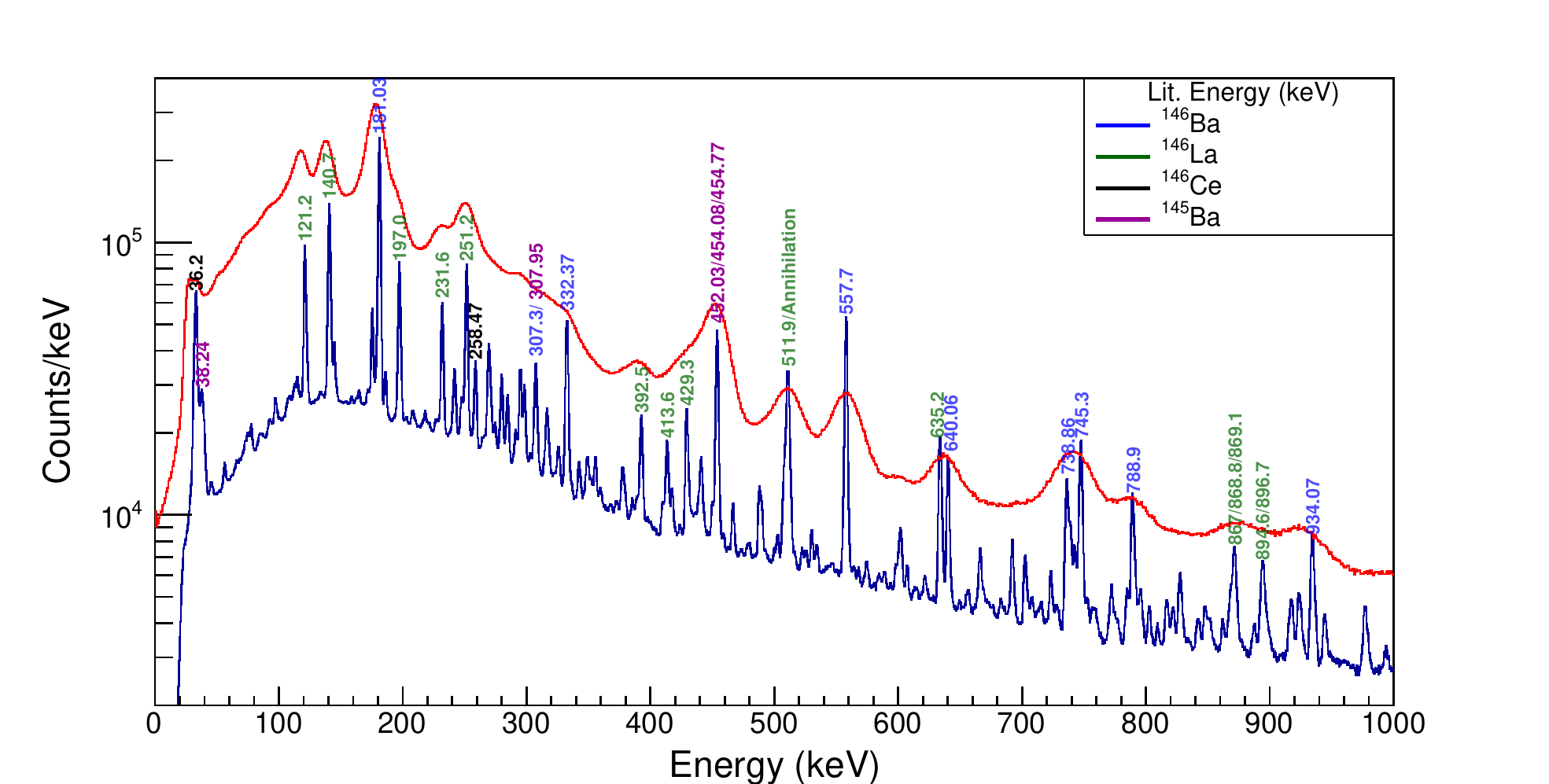}
  \caption{\textit{(Color online)} Energy spectra from the $A$=146 beam. The $\beta$-gated LaBr$_3$(Ce) energy spectrum was generated with an additional optional coincidence in the HPGe array. The red line corresponds to the LaBr$_3$(Ce) array while the blue line shows the HPGe optional coincidence spectrum. Some of the most intense peaks have been identified. See legend for isotope identification.}
  \label{fig:A=146-energy}
\end{figure*}

The tape movement was optimized to enhance the Cs to Ba decay and minimize the activity of the subsequent decay products. The typical tape cycles for the $A=145$ beam were: 0.5\,s of room background collection, 2\,s of beam implantation, 2\,s of beam off to record the decay and 1.5\,s of tape movement without data collection. In the case of the $A$=146 beam, the typical cycles were: 0.5\,s background, 0.4\,s beam on, 1\,s beam off and 1.5\,s of tape movement. The maximum measured beam intensity was $1.8\cdot10^8$\,pps for $^{145}$Cs and $2.7\cdot10^7$\,pps for $^{146}$Cs. However, for both masses an attenuator was placed in the beam line and the mass separator slits run closer than standard in order to reduce the beam intensities down to $3 - 6\cdot10^5$\,pps. The counting rates of the LaBr$_3$(Ce) crystals were $\sim 4$~kHz each. Data obtained with the $A$=146 beam were collected for nearly 50\,hours and for the $A$=145 beam for 10\,hours. No significant contaminants were observed in either beam (see, for example, Fig.~\ref{fig:A=146-energy}).

Energy and timing signals were collected from each detector using the GRIFFIN digital data acquisition system~\cite{Garnsworthy2017}, operated in a triggerless mode. In addition, the anode signals from the LaBr$_3$(Ce) photomultiplier tubes (PMT) were used as input to a set of NIM analogue electronics for fast coincident timing. An Ortec 935 constant-fraction discriminator for each detector fed a set of logic modules that ultimately present the STOP signal to a set of Ortec 566 time-to-amplitude converter (TAC) NIM module for which the output is digitized in a GRIF-16 digitizer. The START of said TAC module was always the ZDS detector. Energy and efficiency were calibrated using standard radioactive sources of $^{133}$Ba, $^{152}$Eu, $^{60}$Co and $^{56}$Co with the necessary corrections for coincidence summing applied.

\begin{figure}
\includegraphics[width=0.99\columnwidth, keepaspectratio]{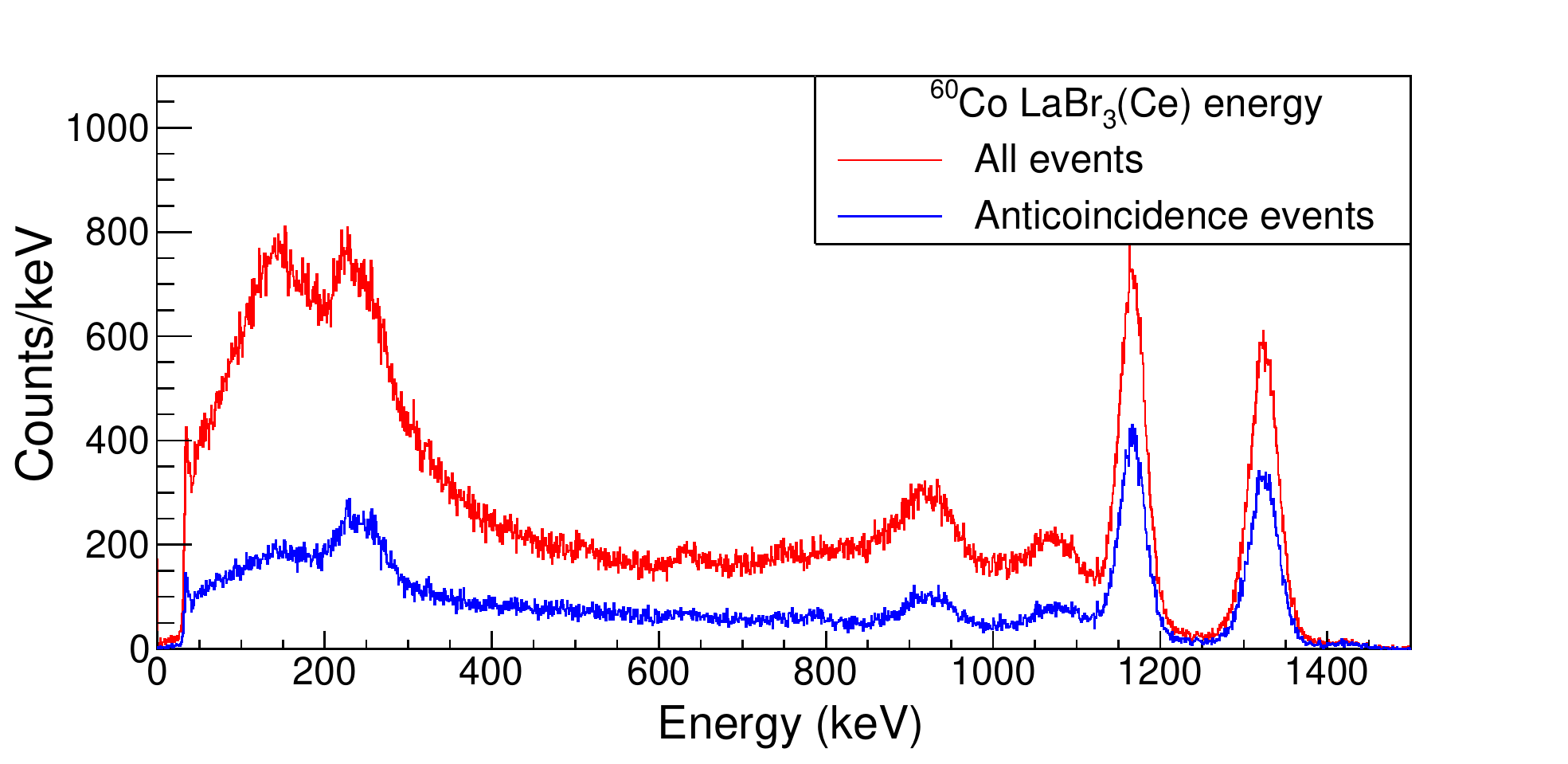}
  \caption{\textit{(Color online)} Energy spectra from a $^{60}$Co offline source taken in LaBr$_3$(Ce)-LaBr$_3$(Ce)-TAC-(HPGe) coincidences. In red, all events independently of the presence of a HPGe coincidence or not. In blue, only events in which there was no coincidence with the GRIFFIN array (anti-coincidence). See text for details.}
  \label{fig:60Co-energy}
\end{figure}

The lifetimes were extracted by the time difference between ZDS and the LaBr$_3$(Ce) crystals as given by the TAC modules. When the decay proceeded via a cascade of two or more $\gamma$ rays, an additional coincidence condition was imposed in the HPGe detectors in order to reduce contributions from other levels feeding the target one. When this was not possible, anti-coincidence with the HPGe detectors were imposed, effectively using the GRIFFIN array as an active Compton-suppression shield, in order to greatly reduce the background contribution in the TAC spectrum. Figure~\ref{fig:60Co-energy} shows this effect with an offline source where it is clearly seen that the contributions to the spectrum which result from Compton scattering are suppressed. In the case of $^{60}$Co, the peak-to-total ratio is improved by over $80\%$, at the cost of losing $\sim 40\%$ of the true coincidences. This loss is caused by the very different geometrical arrangement of the GRIFFIN array (surrounding the source location at primarily backwards angles with respect to the scattering event in the LaBr$_3$(Ce) crystal) in contrast to a standard Compton-suppression shield (surrounding the side and rear faces of the detector primarily at the forward angles of the scattering event). These two numbers have a strong dependence on the multiplicity of the $\gamma$-ray cascade and the energies of the transitions involved.

If the lifetime of the parent level of the decaying transition is long enough (relative to the timing resolution of the system, FWHM$\sim100$~ps), the TAC spectrum will show as the convolution of a prompt Gaussian component plus an exponential decay~\cite{Mach1989, Moszynski1989}. Specifics on the electronic fast-timing technique applied to GRIFFIN can be found in Refs.~\cite{Garnsworthy2019, Olaizola2019}. The whole TAC spectrum can be fitted using an equation of the form:

\begin{equation}
  F(t_j) =\gamma \int_A^{+\infty} e^{- \delta (t_j - t)^2} e^{-\lambda (t - A)}dt \label{eq:convolution_method}
\end{equation}
 
\noindent where the number of counts recorded at time $t_j$ are due to events that happened at time $t$, displaced due to the time jitter. The fitted parameters are: $\gamma$ (normalization factor), $\delta$ (related to the width of the Gaussian prompt distribution), $\lambda$ (decay constant of the parent level of the decaying transition) and $A$ (the centroid of said Gaussian, which is related to the position of a prompt transition of the same energy). When more than one lifetime is present in the spectrum, additional decay components can be added to the fit. 

The use of $\beta-\gamma$ time differences, as opposed to $\gamma-\gamma$ ones, offers an improvement in timing resolution of nearly a factor of 3, mainly because of the reduced volume of ZDS (for the timing performance of thin plastic scintillators, see Ref.~\cite{Fraile2011}). The larger solid angular coverage and its superior intrinsic detection efficiency for charged particles of ZDS versus the LaBr$_3$(Ce) array also ensures the collection of greater statistics. The drawback of this technique is the loss of selectivity. Due to the use of the $\beta$ particle as the START of the TAC, the resulting timing spectrum will present contributions from all the lifetimes of the levels populating directly or indirectly the level of interest. Setting an additional gate on the GRIFFIN array minimizes this effect. Nevertheless, special care was taken when assigning the observed lifetimes to specific levels. For $\gamma$ cascades, the analysis would commence with the transition depopulating the highest-energy level and proceed downwards. Any significantly long lifetime (T$_{1/2} > 100$~ps) found was added as a fixed component in the fit for lower energy levels. It is worth mentioning that the \textit{Pandemonium} effect was not accounted for in the analysis. While this has the potential to have a significant effect on the direct $\beta$ population, it would be necessary for an extremely high number of those weak unobserved transitions to proceed through levels with lifetimes in the hundreds of picoseconds, or longer, range in order to change the result. This is an unlikely scenario.

\section{\label{sec:results}Results}

The lifetimes measured in this work are presented in Table~\ref{tab:lifetimes}. Lifetimes of levels in the even-even $^{144,146}$Ba had been already measured a number of times~\cite{Pritychenko2016} and there is very good agreement with the results of this experiment. While this is a strong validation of the experimental setup and technique, it is worth mentioning that the most precise previous measurements of those isotopes were also performed using an electronic fast-timing method very similar (although simpler) to the one employed here~\cite{Mach1990}. The precision in most of these lifetimes has been slightly increased, mainly due to superior statistics. The exception is T$_{1/2} (2^+_1)$ in $^{144}$Ba; this state was populated in the $\beta$-n branch of $^{145}$Cs decay and hence has much lower statistics.

\begin{table*}
\caption{Summary of the lifetimes measured in this experiment. $E_\text{level}$ - Energy of the level of interest for which the lifetime was measured. $E^\gamma_{decay}$- Transition decaying from level of interest selected in the LaBr$_{3}$(Ce) crystals. $I^\pi_{di}$ and $I^\pi_{df}$ - Spin and parity of the initial and final levels connected by the decaying transition. $E^\gamma_{HPGe}$ - Additional coincidence condition applied on the HPGe crystals. If no HPGe gate is listed, it indicates that the anti-coincidence events were utilized. $I^\pi_{gi}$ and $I^\pi_{gf}$ - Spin and parity of the initial and final levels connected by the transition gated in the HPGe detectors. T$_{1/2}$ - Lifetime measured in this work with the 1 $\sigma$ uncertainties listed in parentheses.  Lit. T$_{1/2}$ - Previous measurements of the lifetime. Individual references are indicated next to the value. All energies and most $I^\pi$ are taken from the evaluations~\cite{Data144, Data145, Data146}. See tables~\ref{tab:transition-strength-145} and~\ref{tab:transition-strength-146} for a list of $I^\pi$ reassigned in this work.  
\label{tab:lifetimes}} 
\begin{center}
\begin{tabular}{cccccccccc}
\hline
\hline

{\bf Isotope} & {\bf $E_{level}$} & {\bf $E^\gamma_{decay}$} & {\bf $I^\pi_{di}$} & {\bf $I^\pi_{df}$} & {\bf $E^\gamma_{HPGe}$} & {\bf $I^\pi_{gi}$} & {\bf $I^\pi_{gf}$} & {\bf T$_{1/2}$} & {\bf Lit. T$_{1/2}$} \\

      & (keV) & (keV) &       &       & (keV) &       &       &  (ps) &  (ps)  \\
\hline
$^{144}$Ba & 199.4 & 199.326 & 2$^{+}_{1}$ & 0$^{+}_{1}$ &  - &  - &  - & 740(90) & 710(20)~\cite{Data144}  \\

\hline
$^{145}$Ba & 112.64 & 112.46 & (7/2)$^{-}_{1}$ &5/2$^{-}_{1}$ & 86.26&(5/2)$^{-}_{2}$&(7/2)$^{-}_{1}$&241(13)&215(12)~\footnote{This value is quoted in Ref.~\cite{Smith1999} citing an unpublished paper. Nevertheless, the values are in reasonable agreement.}  \\
& 175.28 & 175.36 & $(3/2^{-}_{1})$ & 5/2$^{-}_{1}$ & 240.97 & $(5/2,7/2^{-})$ & (3/2)$^{-}_{1}$ & 149(9) &     - \\
& 198.7 & 198.9 & $(5/2_2)^-$ & 5/2$^{-}_{1}$ & 255.9 & $(5/2^{+})$ & $(5/2^-_2)$ & 161(10) &     - \\
& 277.28 & 164.64 & (9/2)$^{-}_{1}$ & (7/2)$^{-}_{1}$ & 112.46 & (7/2)$^{-}_{1}$ & 5/2$^{-}_{1}$ & 172(8) &     -  \\

 & 319.72 & 207.12 & (5/2)$_{1}^{+}$ & (7/2)$_{1}^{-}$ & 112.46 & (7/2)$^{-}_{1}$ & 5/2$_{1}^{-}$ & 124(4) &     -  \\

 & 416.46 & 240.97 & $(5/2,7/2^{-})$ & (3/2)$^{-}_{1}$ & 175.36 & (1/2)$^{-}_{1}$ & 5/2$^{-}_{1}$ & 119(3) &     -  \\

& 435.69 & 435.63 & $(5/2^{+})$ & 5/2$^{-}_{1}$ &  - &  - &  - & $<$100\footnote{An upper limit has been assigned because the lifetime is lower than the range offered by the fitting technique. See text for additional details.} &     -  \\

 & 454.63 & 454.77 & $(5/2^{+})$ & 5/2$^{-}_{1}$ &  - &  - &  - & $<$100\footnotemark[2] &     -  \\

 & 492.12 & 492.08 &  $(5/2,7/2^{-})$ & 5/2$^{-}_{1}$ &  - &  - &  - & $<$100\footnotemark[2] &     -  \\

& 547.09 & 547.06 & (5/2)$^{+}$ & 5/2$^{-}_{1}$ &  - &  - &  - & $<$100\footnotemark[2] &     -  \\

\hline
$^{145}$La &  65.9 &  65.9 &   $(7/2^{+}_{1})$ &   (5/2)$^{+}_{1}$ &  - &   - & - & - &     $9(2)\cdot10^3$~\cite{Clarck1974} \\

 &  96.6 &  96.6 &   $(3/2,5/2,7/2^+)$ &   (5/2)$^{+}_{1}$ &  417.8 &    &  $(3/2,5/2,7/2^+)$ & $18(2)\cdot10^2$ &     -  \\

 & 351.5 & 162.3 &    $(3/2,5/2,7/2^+)$ &    $(3/2,5/2,7/2^+)$ &  91.9 &   $(3/2,5/2,7/2^+)$ &   $(3/2,5/2,7/2^+)$ & $<$100\footnotemark[2] &     -  \\

& 475.3 & 378.8 &  $(1/2-9/2)$ &  $(3/2,5/2,7/2^+)$ &  96.6 &   $(3/2,5/2,7/2^+)$ & (5/2)$^{+}_{1}$ & $<$100\footnotemark[2] &     -  \\

& 492.2 & 303.2 & - &  $(3/2,5/2,7/2^+)$ &  91.9 &  $(3/2,5/2,7/2^+)$ &  $(3/2,5/2,7/2^+)$ & $<$100\footnotemark[2] &     -  \\

 & 514.2 & 417.8 & - &   $(3/2,5/2,7/2^+)$ &  96.6 &  $(3/2,5/2,7/2^+)$ & (5/2)$^{+}_{1}$ & $<$100\footnotemark[2] &     -  \\

\hline
$^{146}$Ba & 181.06 & 181.03 & 2$^{+}_{1}$ & 0$^{+}_{1}$ & 557.7 & 1$^{-}_{1}$ & 2$^{+}_{1}$ & 853(16) & 859(29)~\cite{Mach1990}  \\

& 513.55 & 332.37 & 4$^{+}_{1}$ & 2$^{+}_{1}$ & 181.03 & 2$^{+}_{1}$ & 0$^{+}_{1}$ & $<$100\footnotemark[2] & 0.31(5) ~\cite{Bucher2017}  \\

& 738.79 & 557.7 & 1$^{-}_{1}$ & 2$^{+}_{1}$ & 181.03 & 2$^{+}_{1}$ & 0$^{+}_{1}$ & 159(2) & 160(10)~\cite{Mach1990}  \\

& 820.98 & 640.06 & 3$^{-}_{1}$ & 2$^{+}_{1}$ & 181.03 & 2$^{+}_{1}$ & 0$^{+}_{1}$ & 236(5) & 237(8)~\cite{Mach1990}  \\

 & 1052.38 & 871.49 & 0$^{+}_{2}$ & 2$^{+}_{1}$ & 181.03 & 2$^{+}_{1}$ & 0$^{+}_{1}$ & $<$100\footnotemark[2] & $\leq$26~\cite{Mach1990}  \\

 & 1115.22 & 934.07 & (1,2$^{+}$) & 2$^{+}_{1}$ & 181.03 & 2$^{+}_{1}$ & 0$^{+}_{1}$ & $<$100\footnotemark[2] &     -  \\

\hline
$^{146}$La & 121.16 & 121.2 & $(1^-,2^-,3^+)$ & $(2^{-}_{1})$ & 251.2 & 1$^{+}_{1}$ & $(1^-,2^-,3^+)$ & $<$100\footnotemark[2] &     -  \\

& 140.85 & 140.7 & $(2^+)$ & $(2^{-}_{1})$ & 231.6 & 1$^{+}_{1}$ & $(2^+)$ & $<$100\footnotemark[2] &     -  \\

& 144.62 & 144.7 & $(3^{-})$ & $(2^{-}_{1})$ & 284.5 & 2$^{-}_{4}$ & (3)$^{-}_{1}$ & 1057(16) &     -  \\

& 197.03 &   197.0 & $(1^{-}_{2})$ & $(2^{-}_{1})$ & 175.3 & 1$^{+}_{1}$ & (1)$^{-}_{2}$ & $<$100\footnotemark[2] &     -  \\

& 372.53 & 251.2 & 1$^{+}_{1}$ & $(1^-,2^-,3^+)$ & 121.2 & $(1^-,2^-,3^+)$ & $(2^{-}_{1})$ & $<$100\footnotemark[2] &     -  \\

& 392.61 & 392.5 & $(2^{+})$ & $(2^{-}_{1})$ & 316.3 & 1$^{+}$ & $(2^{+})$ & $<$100\footnotemark[2] &     -  \\

& 429.21 & 284.5 & 2$^{-}$ & (3)$^{-}_{1}$ & 144.7 & (3)$^{-}_{1}$ & $(2^{-}_{1})$ & $<$100\footnotemark[2] &     -  \\

& 466.54 & 269.6 & 2$^{+}$ & (1)$^{-}_{2}$ &   197 & (1)$^{-}_{2}$ & $(2^{-}_{1})$ & $<$100\footnotemark[2] &     -  \\

& 574.5 & 433.6 & (1$^{-}_{7}$,2) & $(2^{+})$ & 140.7 & $(2^{+})$ & $(2^{-}_{1})$ & $<$100\footnotemark[2] &     -  \\

& 708.84 & 279.5 & 1$^{+}$ & 2$^{-}$ & 429.3 & 2$^{-}$ & $(2^{-}_{1})$ & $<$100\footnotemark[2] &     -  \\

& 880.24 & 413.6 & 1$^{+}$ & 2$^{+}$ & 269.6 & 2$^{+}$ & (1)$^{-}_{2}$ & $<$100\footnotemark[2] &     -  \\

& 1064.51 &   692 & 1$^{+}$ & 1$^{+}_{1}$ & 251.2 & 1$^{+}_{1}$ & $(1^-,2^-,3^+)$ & $<$100\footnotemark[2] &     -  \\

\hline
\hline
\end{tabular} 
\end{center}
\end{table*}

\begin{figure}
\includegraphics[width=0.99\columnwidth, keepaspectratio]{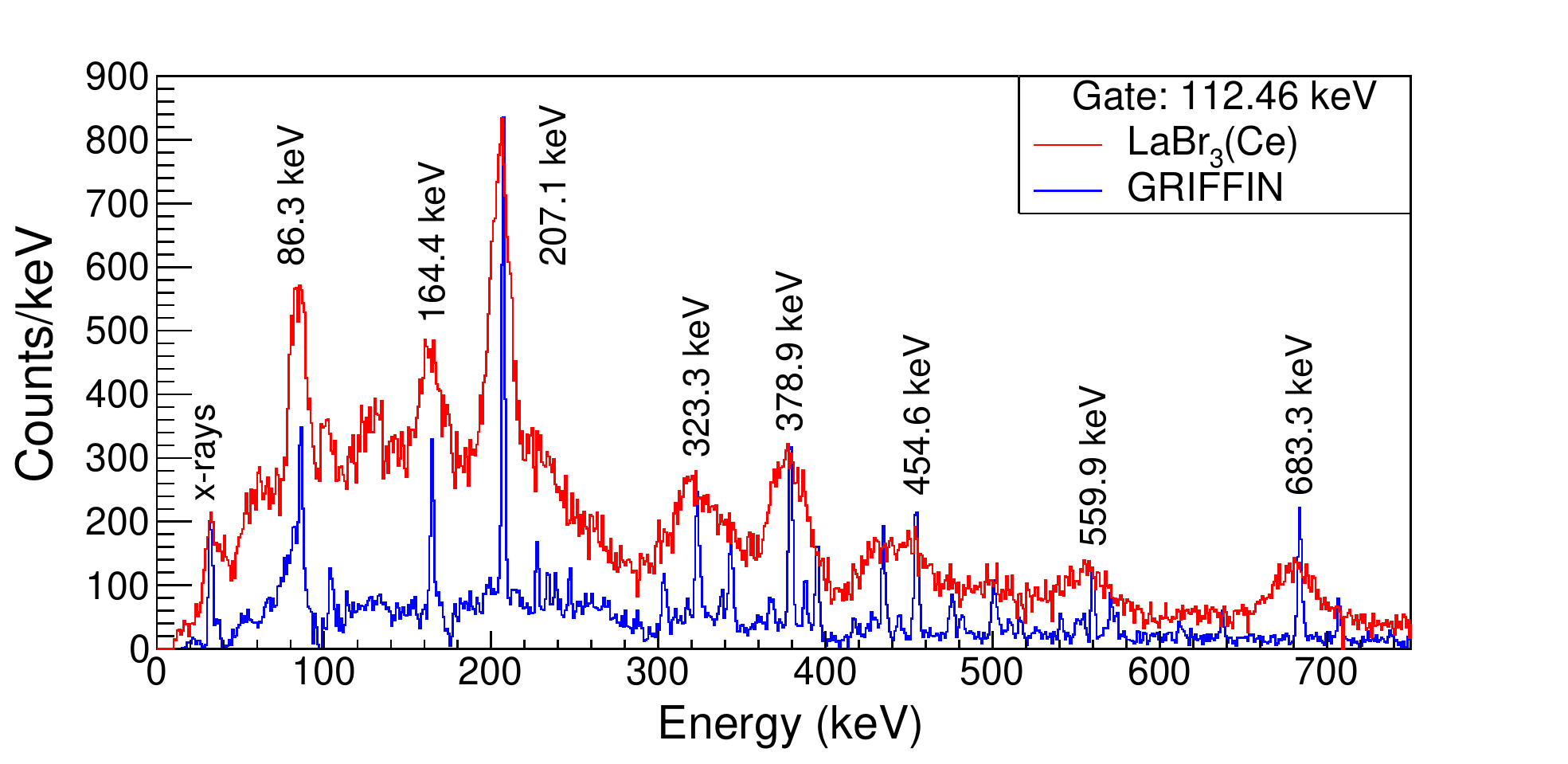}
  \caption{\textit{(Color online)} Energy spectra in $\beta$-LaBr-HPGe coincidence. Red: LaBr$_3$(Ce) energy spectrum with a gate on the 112.46-keV transition detected in GRIFFIN. Blue: GRIFFIN energy spectrum with a gate on the 112.46-keV transition detected in the LaBr array. To better replicate the fast-timing conditions, no time-random events have been subtracted. The most intense transitions are labelled, all belonging to the $^{145}$Cs to $^{145}$Ba decay.}
  \label{fig:Gated-energy-Ba145}
\end{figure}

Figure~\ref{fig:A=146-energy} is an example of the amount of statistics available from the $A$=146 beam in this experiment. The spectrum was generated in $\beta\gamma$($t$) coincidences with an optional hit in the HPGe array. When no hit was recorded in the HPGe, the event was written as an anti-coincidence to be used as an active Compton suppressor. In Table~\ref{tab:lifetimes}, when no HPGe gate is listed, it indicates that the anti-coincidence events were utilized. Figure~\ref{fig:Gated-energy-Ba145} shows energy spectra of the LaBr$_3$(Ce) and HPGe detectors when gating on the 112.46-keV $(7^{}_{}/2^{}_1)^- \rightarrow 5^{}_{}/2^-_1$ transition in $^{145}$Ba. These spectra demonstrate the effectiveness of requiring a coincidence of a specific $\gamma$ ray in either detector system on the cleanliness of the resultant spectra, despite the inferior energy resolution of LaBr$_3$(Ce) crystals compared to the HPGe detectors. The corresponding timing spectra will have a comparable cleanliness.

\begin{figure*}
\includegraphics[width=0.93\textwidth]{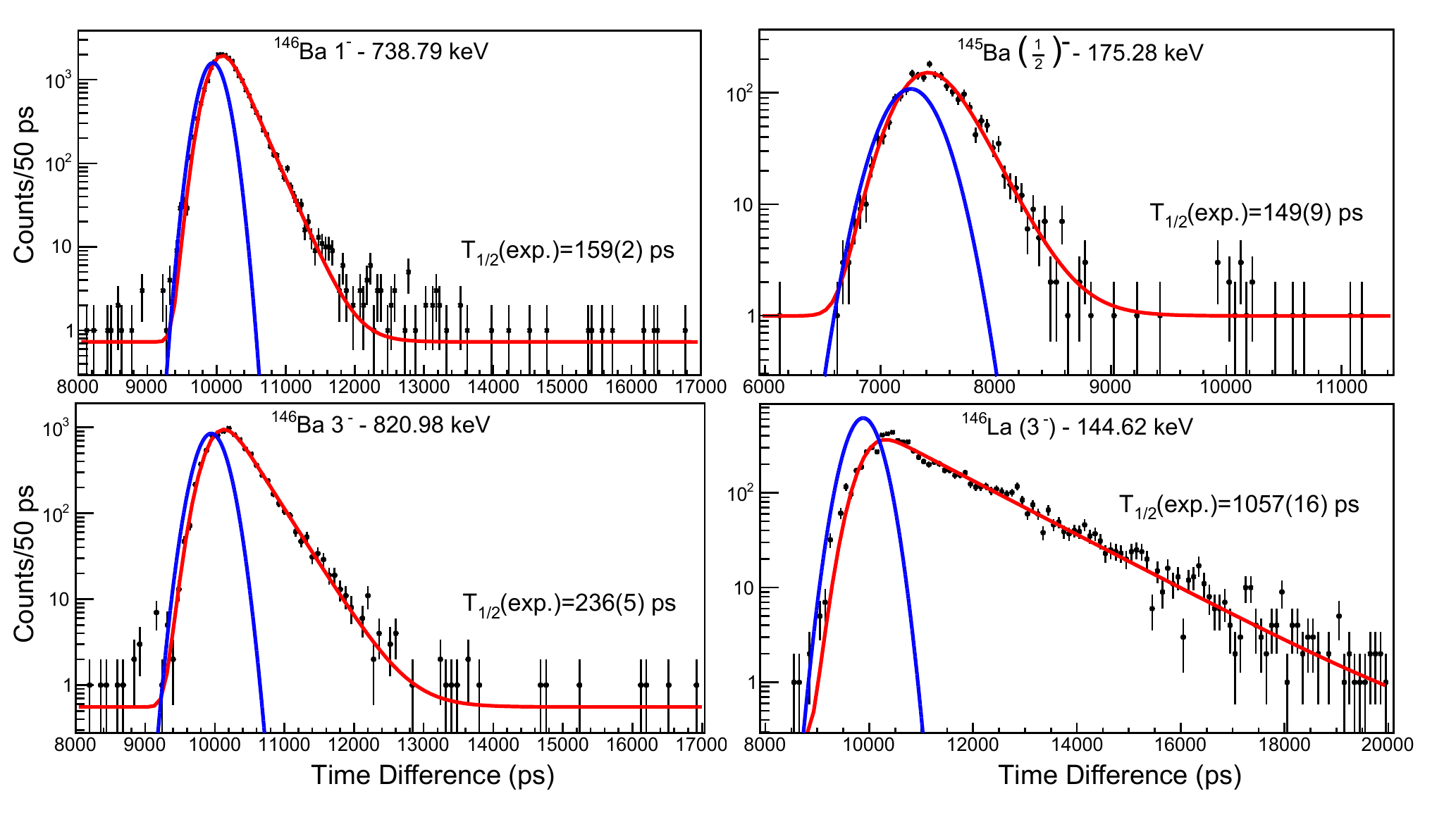}
  \caption{\textit{(Color online)} Example of some of the lifetimes measured in this work using the convolution method. The examples cover the range of lifetimes measured in this experiment. The blue represents the Gaussian prompt component and the red line the convolution of said Gaussian with the exponential decay. The fit was done using the form given in  Eq.~\ref{eq:convolution_method} plus a constant background.}
  \label{fig:lifetimes}
\end{figure*}

Figure~\ref{fig:lifetimes} shows examples of the lifetimes measured in this work for the $A$=145,146 isotopes. The fits were performed using Eq.~\ref{eq:convolution_method} plus a constant background. The blue line corresponds to the Gaussian prompt component and the red line the total fit that includes the exponential decay from which the lifetime is extracted. In each case the fit was repeated by varying the energy gates and the time spectra compression factor. The fitting range was also varied to include more or less time-background, but that was found to have a small impact on the final result. The tail of the Gaussian prompt extends well into the exponential decay component and thus it cannot be neglected in the fitting of all but the longest lifetimes. As such, the so-called chop analysis (repeating the fit to the delayed component with different time intervals in order to study variations in the slope) was not possible and therefore all the fits performed here included the whole time distribution, as shown in Fig.~\ref{fig:lifetimes}. The reported uncertainty is a combination of the statistical error and a systematic error obtained from the variation observed in the tests previously described.  

For short lifetimes the TAC spectrum showed no asymmetry, \textit{i.e.} the delayed component was not long enough for a slope to be fitted in the convoluted timing spectrum. In these cases, a conservative upper limit of T$_{1/2}<100$~ps was assigned to those levels. This value was chosen from the FWHM of the timing resolution. The timing resolution has a dependency with the energy deposited in the scintillators. The observation, or not, of the delayed component also depends on the level of statistics and the peak-to-background ratio of the gated transition. A careful characterization of the timing response of the system would allow slightly more stringent limits to be assigned in these short lifetimes. However, as this was an initial implementation of the fast-timing setup at GRIFFIN, it was not fully optimized or characterized at the time of this data collection. For this reason, the more conservative upper limit of 100~ps has been used in this work for all levels regardless of the transition energy.

Reduced transition probabilities were calculated using the lifetimes from this work presented in Table~\ref{tab:lifetimes} and the evaluated energies, branching ratios and assumed multipolarities based on the $I^\pi$ values (when available, note that in several cases different $I$ have been proposed) provided in Refs.~\cite{Data145, Data146}. Table~\ref{tab:transition-strength-145} presents the calculated values for $^{145}$Ba and $^{145}$La and Table~\ref{tab:transition-strength-146} for $^{146}$La. The values for $^{144,146}$Ba were not calculated again since there are no significant differences from previous works.

\begin{table*}
\caption{Calculated transition strengths for $A$=145 using the lifetimes from this work. Energies and branching ratios were extracted from the NNDC compilations~\cite{Data145}. All transitions have been corrected by their \textit{BrIcc} internal conversion coefficient~\cite{BRICC}. Unless a mixing ratio $\delta$ has been previously measured, pure multipolarity were considered in the extraction of the B(XL). In most cases, a variety of multipolarities has been calculated to facilitate the discussion. All the B(M2) values and higher multipolarities largely exceeded the RUL and thus were not included in the table. \label{tab:transition-strength-145}} 
\begin{center}
\begin{tabular}{ccccccccc}
\hline
\hline
Isotope	& 	E$_\text{level}$ 	& 	 J$^\pi_i$	& 	 E$_\gamma$	& 	 J$^\pi_f$ 	& 	$\delta$	& 	 B(E1) 	& 	B(M1) 	& 	B(E2)	\\	
	& 	(keV)	& 		& 	(keV)	& 		& 		& 	(W.u.)	& 	(W.u.)	& 	(W.u.)	\\	\hline
$^{145}$Ba	&	112.64	&	$(7/2_1)^-$	&	112.46	&	$5/2^{(-)}_1$	&	$-0.40(9)$~\cite{Rzaca2012}	&		&	$2.5(1)\cdot 10^{-2}$	&	$220^{+90}_{-80}$	\\	
	&	175.28	&	$(3/2^-_1)$~\footnote{Spin and/or parity from this work}	&	175.36	&	$5/2^{(-)}_1$	&	$0.7^{+4}_{-3}$~\cite{NNDC}	&		&	$1.5(5)\cdot 10^{-2}$	&	$1.4^{+9}_{-8}\cdot 10^{2}$	\\	
	&	198.7	&	$(5/2_2)^-$	&	86.3	&	$7/2^-_1$	&		&		&	$9.9(17)\cdot 10^{-3}$	&	$7.5(13)\cdot 10^{2}$	\\	
	&		&		&	198.9	&	$5/2^{(-)}_1$	&		&		&	$8.2(6)\cdot 10^{-3}$	&	$1.2(1)\cdot 10^{2}$	\\	
	&	277.28	&	$9/2^-_1$	&	164.64	&	$7/2^-_1$	&	-0.22(7)~\cite{Rzaca2012} 	&		&	$1.8(2)\cdot 10^{-2}$	&	$18(15)$	\\	
	&		&		&	277.12	&	$5/2^{(-)}_1$	&		&		&		&	8.2(5)	\\	
	&	319.72	&	$(5/2)^+$\footnotemark[1]	&	121.01	&	$(5/2_2)^-$	&		&	$1.8(9)\cdot 10^{-5}$	&		&		\\	
	&		&		&	207.12	&	$(7/2_1)^-$	&		&	$1.45(6)\cdot 10^{-4}$	&		&		\\	
	&		&		&	319.84	&	$5/2^{(-)}_1$	&		&	$1.85(8)\cdot 10^{-5}$	&		&		\\	
	&	416.46	&	$(5/2^-, 7/2^-)$\footnotemark[1]	&	240.97	&	$(3/2^-_1)$	&		&		&	$9.8(3)\cdot 10^{-4}$	&	95(3)	\\	
	&		&		&	304.5	&	$(7/2_1)^-$	&		&		&	$4.0(5)\cdot 10^{-4}$	&	2.5(3)	\\	
	&		&		&	416.92	&	$5/2^{(-)}_1$	&		&		&	$3.3(6)\cdot 10^{-4}$	&	1.1(2)	\\	
	&	435.69	&	$(5/2)^+$\footnotemark[1]	&	260.29	&	$(3/2^-_1)$	&		&	$>3.5\cdot 10^{-6}$	&		&		\\	
	&		&		&	323.34	&	$(7/2_1)^-$	&		&	$>1.1\cdot 10^{-5}$	&		&		\\	
	&		&		&	435.63	&	$5/2^{(-)}_1$	&		&	$>2.4\cdot 10^{-5}$	&		&		\\	
	&	454.63	&	$(5/2^+)$\footnotemark[1]	&	38.24\footnote{The 38.24\,keV transition energy is too close to the K electron binding energy (37.4\,keV) for the results to be reliable.}	&	$(5/2^-, 7/2^-)$	&		&	$>3.8\cdot 10^{-3}$	&	$>2.9\cdot 10^{-1}$	&	$>2.0\cdot 10^{4}$	\\	
	&		&		&	255.94	&	$(5/2_2)^-$	&		&	$>1.8\cdot 10^{-5}$	&	$>1.5\cdot 10^{-3}$	&	$>13.6$	\\	
	&		&		&	279.46	&	$(3/2^-_1)$	&		&	$>2.2\cdot 10^{-5}$	&	$>1.7\cdot 10^{-3}$	&	$>12$	\\	
	&		&		&	341.74	&	$(7/2_1)^-$	&		&	$>1.8\cdot 10^{-6}$	&		&	$>0.7$	\\	
	&		&		&	454.77	&	$5/2^{(-)}_1$	&		&	$>1.3\cdot 10^{-5}$	&	$>1.2\cdot 10^{-3}$	&	$>3.4$	\\	
	&	492.12	&	$(5/2^-,7/2^-)$\footnotemark[1]	&	214.52	&	$9/2^-_1$	&		&	$>1.8\cdot 10^{-5}$	&	$>1.5\cdot 10^{-3}$	&	$>$17.6	\\	
	&		&		&	293.2	&	$(5/2_2)^-$	&		&	$>1.3\cdot 10^{-5}$	&	$>1.1\cdot 10^{-3}$	&	$>$7.3	\\	
	&		&		&	492.08	&	$5/2^{(-)}_1$	&		&	$>1.6\cdot 10^{-5}$	&	$>1.4\cdot 10^{-3}$	&	$>$3.4	\\	
	&	547.09	&	$(5/2^+)$	&	227.36	&	$(5/2)^+$	&		&		&	$>3.1\cdot 10^{-3}$	&	$>1.8$	\\	
	&		&		&	348.21	&	$(5/2_2)^-$	&		&	$>1.9\cdot 10^{-6}$	&		&		\\	
	&		&		&	434.71	&	$(7/2_1)^-$	&		&	$>5\cdot 10^{-6}$	&		&		\\	
	&		&		&	547.06	&	$5/2^{(-)}_1$	&		&	$>1\cdot 10^{-5}$	&		&		\\	\hline
$^{145}$La	&	65.9	&	$(7/2^+_1)$	&	65.9	&	$(5/2^+_1)$	&		&		&	$1.9(4)\cdot 10^{-3}$~\footnote{Value extracted using the lifetime measured in~\cite{Clarck1974}}	&	$1.1(2)\cdot 10^2$\footnotemark[1]	\\	
	&	96.6	&	$(3/2^+,5/2^+,7/2^+)$	&	96.6	&	$(5/2^+_1)$	&		&		&	$6.2(7)\cdot 10^{-3}$	&	$2.5(3)\cdot 10^2$	\\	
	&	351.5	&	$(3/2^+,5/2^+,7/2^+)$	&	162.3	&	$(3/2^+,5/2^+,7/2^+)$	&		&		&	$>2.3\cdot 10^{-2}$	&	$>4.7\cdot 10^2$	\\	
	&		&		&	254.9	&	$(3/2^+,5/2^+,7/2^+)$	&		&		&	$>1.9\cdot 10^{-3}$	&	$>$17	\\	
	&		&		&	351.8	&	$(5/2^+_1)$	&		&		&	$>1.3\cdot 10^{-3}$	&	$>$6	\\	
	&	475.3	&	$(1/2-9/2)$	&	286.2	&	$(3/2^+,5/2^+,7/2^+)$	&		&	$>3.4\cdot 10^{-5}$	&	$>5.3\cdot 10^{-3}$	&	$>18$	\\	
	&		&		&	378.8	&	$(3/2^+,5/2^+,7/2^+)$	&		&	$>3.0\cdot 10^{-5}$	&	$>4.7\cdot 10^{-3}$	&	$>10$	\\	
	&	492.2	&		&	303.2	&	$(3/2^+,5/2^+,7/2^+)$	&		&	$>5.2\cdot 10^{-5}$	&	$>4.5\cdot 10^{-3}$	&	$>$28	\\	
	&		&		&	492.7	&	$(5/2^+_1)$	&		&	$>8.2\cdot 10^{-6}$	&	$>7.3\cdot 10^{-4}$	&	$>$1.7	\\	
	&	514.2	&		&	325.2	&	$(3/2^+,5/2^+,7/2^+)$	&		&	$>2.3\cdot 10^{-5}$	&	$>2.0\cdot 10^{-3}$	&	$>11$	\\	
	&		&		&	417.8	&	$(3/2^+,5/2^+,7/2^+)$	&		&	$>2.3\cdot 10^{-5}$	&	$>2.0\cdot 10^{-3}$	&	$>6.5$	\\	\hline

\hline
\hline
\end{tabular} 
\end{center}
\end{table*}

\begin{table*}
\caption{ Same as Table~\ref{tab:transition-strength-145} but for $A$=146. In this case, energies and branching ratios were extracted from the NNDC compilation~\cite{Data146}.\label{tab:transition-strength-146}} 
\begin{center}
\begin{tabular}{ccccccccc}
\hline
\hline
Isotope	& E$_\text{level}$ & J$^\pi_i$ & E$_\gamma$ & J$^\pi_f$ & $\delta$ & B(E1) & B(M1) & B(E2)\\
 & (keV) & & (keV) & & & (W.u.) & (W.u.) & (W.u.)\\ \hline
$^{146}$La	&	121.16	&	$(1^-,2^-,3^+)$	&	121.2	&	$(2_1^-)$	&	+0.04(10)~\cite{Chung1985}	&		&	$>7.6\cdot 10^{-2}$	&	$>0$	\\
	&	140.85	&	$(2_1^+)$	&	140.7	&	$(2_1^-)$	&	-0.66(13)~\cite{Chung1985}\footnote{This experimental $\delta$ is included in order to calculate the hypothetical B(M1) and B(E2) values. It has been assumed $\delta \sim 0$ when calculating the B(E1).}	&	$>5.6\cdot 10^{-4}$	&	$>3.8\cdot 10^{-2}$	&	$>3.8\cdot 10^{2}$	\\
	&	144.62	&	$(3^-)$	&	4	&	$(2_1^+)$	&		&	$1.5(1)\cdot 10^{-2}$	&	$1.18(1)\cdot 10^{-1}$	&	$2.3(1)\cdot 10^{3}$	\\
	&		&		&	144.7	&	$(2_1^-)$	&	+0.61(10)~\cite{Chung1985}	&		&	$2.7(1)\cdot 10^{-3}$	&	27(2)	\\
	&	197.03	&	$(1)$	&	56.4	&	$(2_1^+)$	&		&	$>6.1\cdot 10^{-4}$	&	$>4.6\cdot 10^{-2}$	&	$>5.8\cdot 10^3$	\\
	&		&		&	75.9	&	$(1^-,2^-,3^+)$	&		&		&	$>3.9\cdot 10^{-3}$	&	$>3.7\cdot 10^{2}$	\\
	&		&		&	197	&	$(2_1^-)$	&	-0.10(12)~\cite{Chung1985}	&		&	$>2.1\cdot 10^{-2}$	&	$>3.1$	\\
	&	372.53	&	$1^+_1$	&	77.7	&	$(2)$	&		&	$>1.2\cdot 10^{-4}$	&	$>1.1\cdot 10^{-2}$	&	$>9.4\cdot 10^2$	\\
	&		&		&	175.3	&	$(1)$	&		&	$>5.7\cdot 10^{-5}$	&		&		\\
	&		&		&	231.6	&	$(2^+_1)$	&	0.39(6)	&	$>5.5\cdot 10^{-5}$	&	$>4.3\cdot 10^{-3}$	&	$>6.9$	\\
	&		&		&	251.5	&	$(1^-,2^-,3^+)$	&		&	$>7.8\cdot 10^{-5}$	&		&		\\
	&		&		&	372.5	&	$(2_1^-)$	&		&	$>1.2\cdot 10^{-6}$	&		&		\\
	&	392.6	&	$(2^+)$	&	392.5	&	$(2_1^-)$	&		&	$>4.0\cdot 10^{-5}$	&		&		\\
	&	429.17	&	$2^-$	&	284.5	&	$(3^-)$	&	+0.39(25)~\cite{Chung1985}	&		&	$>2.5\cdot 10^{-3}$	&	$>2.6$	\\
	&		&		&	429.3	&	$(2_1^-)$	&	+0.66(11)~\cite{Chung1985}	&		&	$>1.3\cdot 10^{-3}$	&	$>1.7$	\\
	&	466.54	&	$(2)$	&	94	&	$1^+_1$	&		&		&	$>4.7\cdot 10^{-3}$	&	$>290$	\\
	&		&		&	139.8	&	$(3)$	&		&	$>1.1\cdot 10^{-4}$	&	$>1.0\cdot 10^{-4}$	&	$>2.9\cdot 10^2$	\\
	&		&		&	171.6	&	$(2)$	&		&	$>2.6\cdot 10^{-5}$	&	$>2.3\cdot 10^{-3}$	&	$>44$	\\
	&		&		&	269.6	&	$(1)$	&		&	$>7.3\cdot 10^{-5}$	&		&		\\
	&		&		&	466.8	&	$(2_1^-)$	&		&	$>2.8\cdot 10^{-6}$	&		&		\\
	&	574.5	&	$(1^-,2)$	&	107.9	&	$(2)$	&		&	$>4.8\cdot 10^{-5}$	&	$>2.7\cdot 10^{-3}$	&	$>1.3\cdot 10^2$	\\
	&		&		&	145.3	&	$2^-$	&		&	$>4.7\cdot 10^{-5}$	&	$>3.3\cdot 10^{-3}$	&	$>80$	\\
	&		&		&	164.6	&	$(3)$	&		&	$>5.5\cdot 10^{-5}$	&	$>4.2\cdot 10^{-3}$	&	$>81$	\\
	&		&		&	247.8	&	$(3)$	&		&	$>2.9\cdot 10^{-5}$	&	$>2.4\cdot 10^{-3}$	&	$>22$	\\
	&		&		&	279.5	&	$(2)$	&		&	$>9\cdot 10^{-6}$	&	$>7.8\cdot 10^{-3}$	&	$>5.6$	\\
	&		&		&	377.5	&	$(1)$	&		&	$>1.4\cdot 10^{-5}$	&	$>1.2\cdot 10^{-3}$	&	$>4.8$	\\
	&		&		&	433.6	&	$(2_1^+)$	&		&	$>1.4\cdot 10^{-6}$	&	$>1.3\cdot 10^{-4}$	&	$>3.9\cdot 10^{-1}$	\\
	&		&		&	574.5	&	$(2_1^-)$	&		&	$>2.3\cdot 10^{-6}$	&	$>2.1\cdot 10^{-4}$	&	$>3.5\cdot 10^{-1}$	\\
	&	708.79	&	$1^+$	&	279.5	&	$2^-$	&		&	$>5.0\cdot 10^{-5}$	&		&		\\
	&		&		&	291.5	&	$(2)$	&		&	$>1.1\cdot 10^{-5}$	&	$>9.6\cdot 10^{-4}$	&	$>6.4\cdot 10^{-1}$	\\
	&		&		&	316.3	&	$(1^-,2^-,3^+)$	&		&		&	$>2.1\cdot 10^{-3}$	&	$>12$	\\
	&		&		&	335.8	&	$1^+_1$	&		&		&	$>1.7\cdot 10^{-4}$	&	$>8.4\cdot 10^{-1}$	\\
	&		&		&	511.9	&	$(1)$	&		&	$>9.6\cdot 10^{-7}$	&		&		\\
	&		&		&	568.2	&	$(2_1^+)$	&		&	$>2.4\cdot 10^{-7}$	&		&		\\

\hline
\hline
\end{tabular} 
\end{center}
\end{table*}

\section{\label{sec:discussion}Discussion}

The calculated reduced transition probabilities [B(XL)] shown in Tables~\ref{tab:transition-strength-145} and~\ref{tab:transition-strength-146} were used to deduce information on the nuclear structure of low-lying states of the studied isotopes. To do so, we considered the recommended upper limits (RUL) for $90<A<150$ from the compilation~\cite{Endt1981}: B(E1)$<10^{-2}$, B(M1)$<1$, B(E2)$<300$ and B(M2)$<1$~W.u. As this compilation is from 1981, there is a concern that these RUL are now out dated. However, a survey of the evaluated data available for $A$=140-150 did not identify any new B(E2) values above the 300~W.u. RUL.

The implications of the new lifetime data are discussed in the following subsections for each isotope in turn. These discussions make use of the lifetimes measured in this work in combination with information extracted from other experiments using $\gamma \gamma$ angular correlations, conversion electron spectroscopy or $\beta$ decay, as referenced throughout the text.

\subsection*{$^{145}$Ba \label{sec:Ba145}}

The g.s. of $^{145}$Ba was evaluated to have firm $I^\pi = 5/2^{-}$ in the ENSDF compilation~\cite{Data145}. This ground state spin was firmly established as $I=5/2$ by a collinear fast-beam laser spectroscopy~\cite{Wendt1988} experiment. However, it should be noted that the technique can only establish $I$, not the parity of the state, and thus in the nuclear moments compilation~\cite{Stone2005} it appears as tentative $5/2^{(-)}$, based on the systematics of $N$=89 isotones. This tentative assignment will be used in this discussion. 

\begin{figure*}
\includegraphics[width=0.99\textwidth, keepaspectratio]{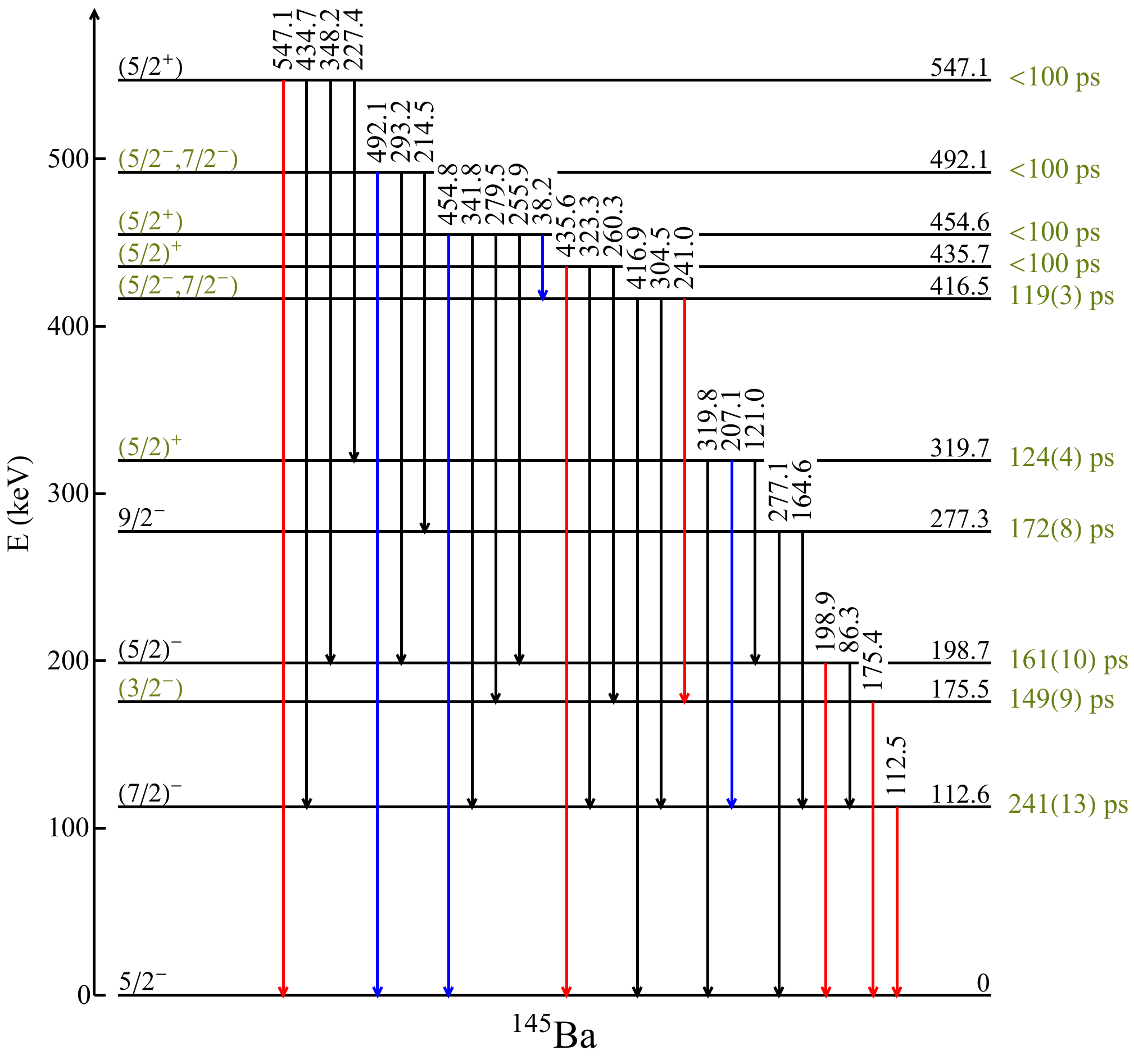}
  \caption{\textit{(Color online)} Partial level scheme of $^{145}$Ba populated in the $\beta^-$ decay of $^{145}$Cs. The energies, intensities and spin-parities in black of levels and trasitions are taken from the evaluation~\cite{Data145}. Following the NNDC color scheme, red transitions have $<10\%$ of the intensity of the strongest transition (112.5-keV one in this case), blue transitions have $2\%<I_\text{max}<10\%$ and black transitions $I_\text{max}<2\%$. The lifetimes to the right of the energy levels and the spin-parities to the left in green are from this work. See text for additional details on the assignment of spin-parities.}
  \label{fig:145Ba-scheme}
\end{figure*}

A partial level scheme is presented in Fig.~\ref{fig:145Ba-scheme} to guide the following discussion.

\textbf{The 112.6-keV level $(7/2_1)^-$:} $I^\pi=7/2^-_1$ was assigned as firm in Refs.~\cite{Jones1996, Zhu1999Ba, Rzaca2012}, although it is only based in band membership, and therefore it was evaluated as tentative in Ref.~\cite{Data145}. There are two measured values for the mixing ratio ($\delta$) of the 112.6\,keV transition to the g.s.: $0.13^{+7}_{-6}$ using directional correlations from oriented nuclei (DCO)~\cite{Jones1996} and $-0.40(9)$ from $\gamma-\gamma$ angular correlations~\cite{Rzaca2012}. The latter is in agreement with the $\delta$ value extracted from the conversion electrons coefficients measured in Refs.~\cite{Rzaca2012, Robertson1986, Rapaport1982}, and thus will be adopted in this work. The calculated transition probability yields B(E2)=220(80)~W.u. This is a surprisingly large value, significantly more enhanced than the one its neighbouring Ba isotopes present, with B(E2;$2^+_1\rightarrow0^+_1$)=45.2(25)~W.u. for $^{144}$Ba and B(E2;$2^+_1\rightarrow0^+_1$)=59.1(29)~W.u. for $^{146}$Ba. Although the uncertainty is large and the systematic is incomplete, this B(E2;$7/2_1^- \rightarrow 5/2^(-)_1$) value seems to be also much larger than the other N=89 odd-isotones, which B(E2) decreases as they approach the proton mid-shell to a minimum for $^{155}$Dy of B(E2;$7/2_1^- \rightarrow 5/2^-_1$)=80(24)~W.u. On the other hand, it is remarkable how constant the B(M1) component of these $7/2_1^- \rightarrow 5/2^-_1$ transition is along the $N$=89 chain, at least up to $^{155}$Dy, the heaviest isotone for which this value has been measured. In similarity with the other $N=89$ isotones, this first excited state can be considered part of the g.s. band built on the $\nu5/2[523]$ configuration.

\textbf{The 175.3-keV level \soutthick{$(1/2^-_1)$} $(3/2^-_1)$:} The measured lifetime for this state is surprisingly short. Assuming a pure $E2$ character for the $1/2^-_1 \rightarrow 5/2^(-)_1$ transition (the only branch observed decaying from this state), yields a B(E2) value of 400(20)~W.u. This value is significantly larger than the RUL of BE(2)$<300$~W.u. and is a strong indication that this transition may have a different multipolarity. In order to bring this value in line with the measured B(E2) in neighbouring nuclei, the lifetime would have to be 5 to 10 times longer, which seems very unlikely. Careful checks were made in the analysis to confirm accurate timing for low-energy transitions. A more likely explanation is that the spin assignment of this parent state is incorrect.


The $(1/2^-)$ assignment comes from angular correlations reported in Ref.~\cite{Robertson1986}. The authors claimed to have observed isotropic ($A_{22}=0$ and $A_{44}=0$, where $A_{ii}$ are the normalized coefficients of the Legendre polynomials) angular distributions when the 175~keV transition was used as the gate, a strong suggestion that this level has spin 1/2. However, taking the uncertainties the authors reported as a sensitivity of $|A_{22}|, |A_{44}|<0.05$, a large number of spin combinations and mixing ratios ($\delta_1$ and $\delta_2$) can fit their measured distributions. As a relevant example for this discussion,  angular correlations for a $5/2 \rightarrow 3/2 \rightarrow 5/2$ cascade are almost flat for $\delta_1=\delta_2=0$, with $A_{22}=0.01$ and $A_{44}=0$, or $-0.05<A_{22}<+0.05$ for a wide range of $\delta_1$, $\delta_2$ values. It also should be noted that they show the angular correlation coefficients for two $0 \rightarrow 2 \rightarrow 0$ transitions in $^{142}$Ba and the agreement with theory is poor. This calls into question, at the very least, their reported uncertainties.


Even more surprisingly, the measured conversion coefficient~\cite{Robertson1986} fits better for a M1 character than for an E2, although the difference between both is only 15\% (still, the reported precision of the experiment was significantly higher). It is worth noting the very good agreement for the conversion coefficient with~\cite{Rapaport1982}. These two results add further weight to the reassignment of this transition as M1 with an apparently small admixture of E2.

Lastly, the authors of Ref.~\cite{Robertson1986} also claimed that the 175.4-keV transition to the g.s. was assigned as an E2 in Ref.~\cite{Schussler1981}. However, Ref.~\cite{Schussler1981} describes results on $A$=147, with only one unexplained plot where it shows this transition marked having E2 character and assigning $(3/2^-)$ to the 175-keV level.

In light of all this, the 175-keV state should be reassigned as having spin and parity of $(3/2^-)$. This reassignment is strongly supported by the present lifetime data and favored by the conversion coefficient (the conversion coefficient for E2 would have a 2$\sigma$ deviation). This reassignment is not contradicted by the reported $\gamma-\gamma$ angular correlations and the first-forbidden log(ft)=6.12(7) value~\cite{Data145} neither of which can distinguish between $1/2^-$ or $3/2^-$. 



\textbf{The 198.7-keV level $(5/2_2)^-$:} The measured lifetime discard any significant E2 component for the 86.3- and 198.9-keV transitions, suggesting $I=5/2, 7/2$. The experimental electron conversion coefficient from works~\cite{Robertson1986, Rapaport1982} clearly rejects an E1 character for the 198.9-keV transitions, thus implying the parity of the level is negative. The evaluated direct $\beta$ feeding yields log$(ft)=6.05$, suggesting $I^\pi =(5/2^-_2)$ as the most likely option. This is in agreement with evaluation~\cite{Data145} and suggest a dominant M1 character for both transitions.

\textbf{The 277-keV level $9/2_1^-$:} The spin of $9/2$ was firmly established from data obtained from fission experiments~\cite{Rzaca2012}. The B(M1) and B(E2) strength of the transitions are average. The large error in the B(E2) portion of the $9/2^- \rightarrow 7/2^-$ transition arises from the uncertainty in $\delta$.

\textbf{The 320-keV level \soutthick{$(3/2^+)$}\footnote{After making the NNDC evaluators aware of the issues here discussed, this spin-parity has been updated to $(5/2^+)$ in ENSDF.} $(5/2)^+$:} The spin of the 320-keV level was determined as $(3/2^+)$ from the E1 character of the 207-keV transition to the 112-keV state. This, in turn, was extracted from the measured conversion coefficients, $\alpha_K(E1)=5.3(20)\cdot 10^{-2}$~\cite{Robertson1986}, compared with the theoretical $\alpha_K(E1)=2.6\cdot 10^{-2}$, $\alpha_K(M1)=11\cdot 10^{-2}$ and $\alpha_K(E2)=12\cdot 10^{-2}$. The spin $(3/2^+)$ was assigned when the 112-keV state was wrongly established as a $(3/2,5/2)^-$, hence the error. Now that the spin of the 112-keV state is firmly established as $7/2^-$, the correct assignment based on the conversion coefficient would be $(5/2,7/2,9/2)^+$. The $I^\pi=9/2^+$ possibility can be rejected by the presence of a transition to the $5/2^{(-)}_1$ g.s., which would imply a M2 character orders of magnitude above the RUL. The evaluated log(\textit{ft})=6.99(10)~\cite{Data145} value is high, but still within the limits of observed allowed transitions~\cite{Singh1998}. With the parent g.s. established as $I^\pi=3/2^+$, the $7/2^+$ option can be ruled out and it leaves $(5/2^+)$ as the most likely spin. Since the assignment is partly based on a log(\textit{ft}) value situated on the higher end of the distribution, it cannot be completely ruled out that this transition is a second forbidden and the state $7/2^+$. Thus, the authors think the tentative character of the spin assignment is warranted.

Assuming $I^\pi=5/2^+$, this level is a strong candidate for the parity doublet of the g.s. The intrinsic electric dipole moment, $D_0$, can be extracted from the B(E1) assuming a strong coupling limit and an axial shape of the nucleus:

\begin{equation}
    \text{B(E1)} = \frac{3}{4\pi}D_0^2 \left\langle I_i K_i10 | I_f K_f \right\rangle^2 \label{eq:e_dipole}
\end{equation} 

\noindent where $\left\langle I_i K_i10 | I_f K_f \right\rangle$ are the Clebsch-Gordan coefficient. The $D_0$ for the $5/2^+_1 \rightarrow 7/2^-_1$ and $5/2^+_1 \rightarrow 5/2^(-)_1$ transitions are 0.062(12) and 0.014(3) $e\cdot$fm, respectively. This is in agreement with the indirect estimation of $D_0 \sim 0.05(1) e\cdot$fm for the $19/2^-_1$ state made in Ref.~\cite{Jones1996}. These are small electric dipole moments, closer to the notoriously quenched $D_0$ in $^{146}$Ba than the enhanced values of the other even-Ba isotopes~\cite{Mach1990, Urban1997}.

\textbf{The 416-keV level \soutthick{$(5/2^-)$} $(5/2^-, 7/2^-)$:} The spin-parity of this level was established from $\gamma-\gamma$ angular correlations. However, the resulting positive A$_{44}$ coefficient only indicates a spin larger than $5/2$~\cite{Robertson1986}. The B(M1/E2) values are rather small, although reasonable for non-collective E2 transitions. The obtained value for the 241-keV transition was calculated assuming pure E2 character, but with the reassignment of the 175-keV state as 3/2, some M1 character must be considered. We note that the 241-keV transition was assigned a pure E2 character based on the measured conversion coefficient, but $\alpha$ is almost the same for both M1 and E2 multipolarities, making it impossible to distinguish between them. However, the measured K/L ratio would seem to favour the M1 character. The 241-keV transition is significantly enhanced in comparison to the other two transitions. The observed feeding via an E1 transition from the 454-keV state limits the spin of the 416-keV state to be $7/2^-$ or lower. The same is true for the transition from this level to the $(3/2_1^-)$ state. Considering all the available information, the 416-keV can be assigned to have a spin and parity of $(5/2^-, 7/2^-)$, with the $(5/2^-)$ spin being more likely if the pure M1 character of the 241-keV transition is accepted.

\textbf{The 435-keV level \soutthick{$3/2^+$} $(5/2)^+$:} The spin and parity of the 435-keV state was established in Ref.~\cite{Robertson1986} from the E1 character of the 435-keV transition through the conversion coefficients, but this gives a spin range of $3/2,5/2,7/2^+$. The 7/2 option can be discarded from log(\textit{ft}) considerations. The $3/2^+$ assignment can be discarded by the observation of the 323.3-keV transition to the $(7/2)^-_1$ state. With the upper limit of the lifetime measured in this work, it would yield a B(M2) value orders of magnitude above the RUL. Thus, the most likely assignment is $(5/2)^+$. It should be noted that the authors of Ref.~\cite{Robertson1986} discarded $I^\pi=5/2^+$, claiming that the 260-keV transition to the $1/2^-$, 175-keV state would have M2 character and as such they would have observed it in the electron spectrum. Now that the spin assignment for the 175-keV has been revised to be $(3/2^-)$, this 260-keV transition is in fact an E1 multipolarity with a much smaller internal conversion coefficient and therefore provides an explanation as to why it was not observed.  

\textbf{The 454-keV level \soutthick{$(3/2^-)$} $(5/2^+)$:} There is a lack of evidence to assign this state as $(3/2^-)$. An $\alpha_K$ measurement is reported in Ref.~\cite{Robertson1986}, but there seems to be different typos in that line of the table. The values reported in ENSDF for the 454-keV transition are consistent with an E1 multipolarity, not with a M1 character. With the compiled intensity for the 38-keV transition, the B(XL) it yields are unphysical, B(E2)$>2.0\cdot10^{4}$~W.u. and B(M1)$>2.9\cdot10^{-1}$~W.u. The energy of this transition is only 1~keV away from the K-electron binding energy, thus physical processes such as electron-electron correlations that are not taken into account, could affect the resulting conversion coefficient. Either the intensity or the $\alpha$ coefficient (or a combination of both) must be wrong by at least 1 order of magnitude for the lower limits to be physically plausible (and the E2 component to be highly suppressed).

Assuming the E1 character of the 454-keV transition would imply that all the other transitions decaying from this state have E1(+M2) character also. In this case the 38-keV would present B(E1)$>3.8\cdot10^{-3}$~W.u., in line with the enhanced E1 transition found between parity doublets of octupole deformation. A $3/2^+$ spin would imply that the 341.7-keV transition to the $7/2^-_1$ state has a pure M2 character. With the measured upper limit for the lifetime of this level, that would yield B(M2)$>70$~W.u., far larger than the RUL of 1~W.u. for this mass region~\cite{Endt1981}. A similar argument using the 279.5-keV transition to the $(3/2^-_1)$ state can be used to discard a $7/2^+$ spin. We, thus, suggest a tentative $I^\pi=(5/2^+)$.

\textbf{The 492-keV level $(5/2^-,7/2^-)$:} The lower limits in the B(XL) of all the transitions are within the RUL except for M2 character, that can be safely discarded. Thus, $I^\pi=(5/2^-, 7/2^{+/-}, 9/2^-)$ are possible. The measured direct $\beta$ decay is compatible with either first forbidden (log$(ft)=6.44$) or unique first forbidden (log$(ft)=8.56$), which further limits the spins to $I^\pi=(5/2^-, 7/2^{-})$.

\textbf{The 547-keV level $(5/2^+)$:} The spin parity was assigned by assuming an E1 character for the 547-keV transition. This E1 was assigned in Ref.~\cite{Robertson1986} because any higher order multipolarity would have shown electrons in their spectrum. The spins were further constrained by the evaluator because of the presence of other transitions.

\subsection*{$^{145}$La \label{sec:La145}}

\begin{figure*}
\includegraphics[width=0.99\textwidth, keepaspectratio]{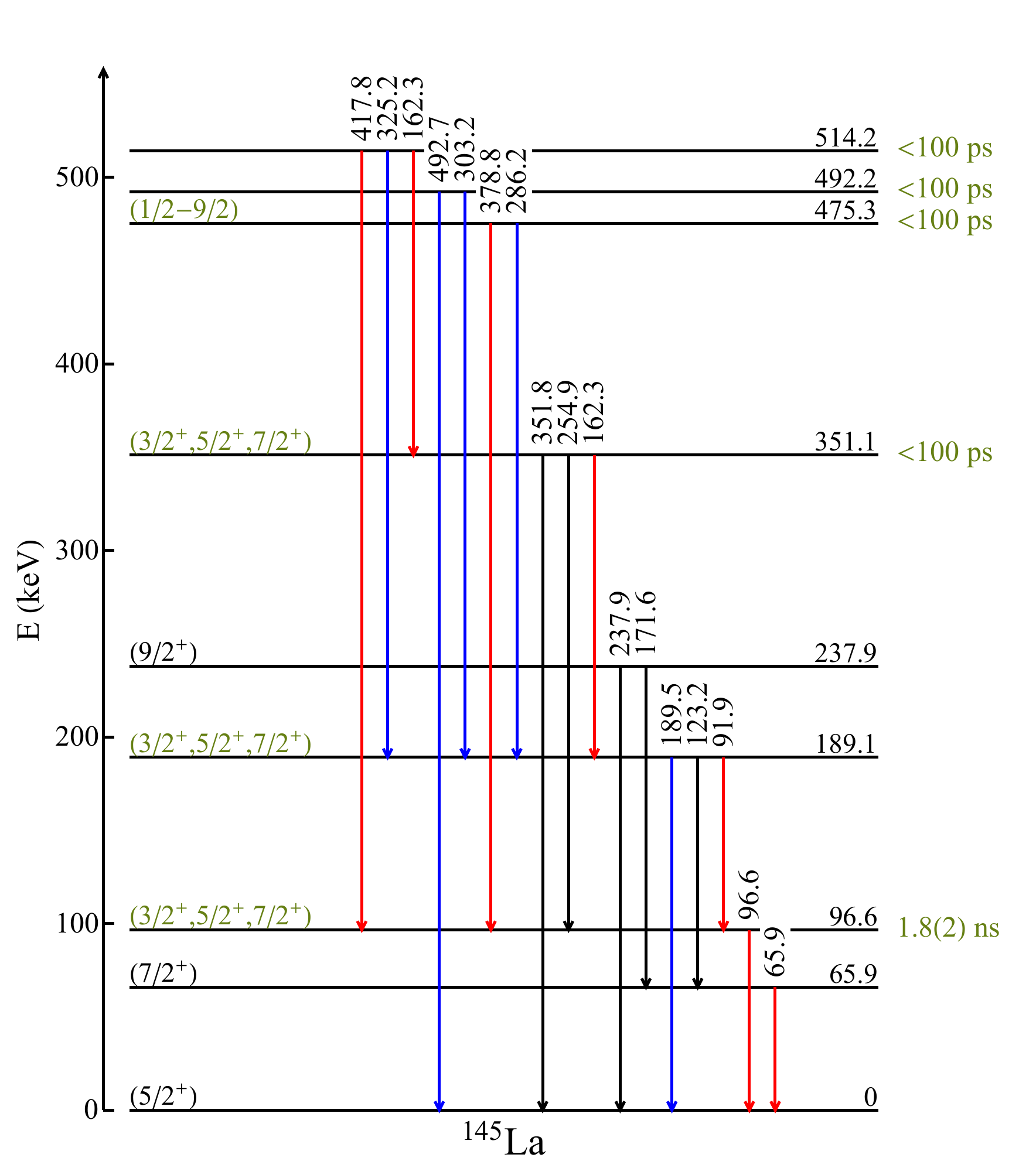}
  \caption{\textit{(Color online)} Same as Fig.~\ref{fig:145Ba-scheme} for the partial level scheme of $^{145}$La populated in the $\beta^-$ decay of $^{145}$Ba~\cite{Data145}.}
  \label{fig:145La-scheme}
\end{figure*}

The ground and first excited states are assigned tentative $(5/2^+)$ and $(7/2^+)$ spin-parities, respectively, in the most recent Evaluation~\cite{Data145}. The authors were not able to access the original arguments for such assignments, but they will be used as the basis for the following discussion. A partial level scheme is presented in Fig.~\ref{fig:145La-scheme}.

\textbf{The 65.9-keV level $(7/2^+_1)$:} Urban \textit{et al.}~\cite{Urban1996} measured a value of $\alpha_K^\text{exp}=4(1)$ as the internal conversion coefficient for this transition. Despite its assignment as a pure M1 ($\alpha_K^\text{BrIcc}=3.07(5)$) in the Evaluation~\cite{Data145}, this experimental value seems to favour an E2 character ($\alpha_K^\text{BrIcc}=3.81(6)$), although the relatively large uncertainty allows for a wide range of mixing ratio with M1. A pure M1 character would not be very collective. However, assuming a pure E2 character, this suggests a high degree of collectivity for this transition and therefore a large deformation for the nucleus.

\textbf{The 96.6-keV level \soutthick{$(^+)$} $(3/2^+,5/2^+,7/2^+)$:} In Ref.~\cite{Data145} the 96.6-keV transition was compiled as a pure M1 multipolarity on the basis of the measured K/L electron ratio. Since this transition feeds the $(5/2^+)$ ground state, this implies $I^\pi=(3/2^+,5/2^+,7/2^+)$ for this level. The calculated transition strengths seem to favour this assignment. A pure E2 transition would yield a large B(E2) value (approaching the RUL B(E2)$<300$~W.u.~\cite{Endt1981}), while the obtained B(M1) seems more in line with other B(M1) values in the region.

The non-observation of a 30.7-keV transition connecting the 96.6- and 65.9-keV states does not help limit the spin of the level. A pure E2 character would suppress the branching ratio to below 0.005 while a pure M1 would need to be below 0.01, in both cases outside the sensitivity of past or present experiments on the nucleus for transitions of this low energy. 

The direct $\beta$-population to the state was measured using a Total Absorption $\gamma$-ray Spectrometer (TAGS)~\cite{Greenwood1997}. The extracted log($ft$) suggests a first forbidden transition from the $5/2^{(-)}_1$ g.s. of $^{145}$Ba, which also suggest the $(3/2^+,5/2^+,7/2^+)$ spin possibilities.

Fission experiments populated the 237.9-keV state with $I^\pi=(9/2^+)$, without observing any transition decaying into this level~\cite{Zhu1999La}. This non-observation could discard the $5/2^+$ and $7/2^+$ assignments, making $3/2^+$ the more likely candidate. Nevertheless, we cannot rule out structure differences between the states that suppressed such a hypothetical transition.

\textbf{The 189.0-keV level \soutthick{$(^+)$} $(3/2^+,5/2^+,7/2^+)$:} The M1 character, with minimal E2 component, of the 91.9-keV transition was firmly established from electron spectroscopy~\cite{Data145}. The log($ft$)=6.6~\cite{Pfeiffer1978} value to this level suggests a first forbidden transition, thus the likely spins are $(3/2^+,5/2^+,7/2^+)$.

\textbf{The 351.5-keV level \soutthick{$(^+)$} $(3/2^+,5/2^+,7/2^+)$:} The 162.3-keV transition was tentatively assigned M1 character from electron spectroscopy~\cite{Data145}. The lifetime upper limit from this work yields a B(E2) lower limit well above the RUL, thus discarding any significant E2 component, in good agreement with the measured conversion coefficient. Once again, log($ft$)=6.3~\cite{Pfeiffer1978} suggests a first forbidden $\beta$ decay to this level and suggests spin and parity of $I^\pi=(3/2^+,5/2^+,7/2^+)$. The other two transitions yield reasonable B(XL) lower limits, that can be interpreted as either pure M1 or E2 or an admixure of both.

\textbf{The 475.3-keV level $(1/2-9/2)$:} The lifetime upper limit does not allow to distinguish if the transitions depopulating this level are of E2, M1 or E1 character, although it discards the possibility of them being of M2 multipolarity. This only limits the spin to be $I\leq9/2$, and does not constrain the parity.

\subsection*{$^{146}$La \label{sec:La146}}

The ground state has a tentative assignment of $I^\pi=(2^-)$~\cite{Data146}, which will be used in the following discussion. The removal of the parenthesis for this state in the online NuData 2.8 has been confirmed to be a typo. Most of the spin parities of the excited states in this nucleus were established in $\gamma-\gamma$ angular correlation measurements involving the 372-keV level. This state has a firmly assigned spin and parity of $I^\pi=1^+$, but solely on the basis of the apparent log(\textit{ft}) from a $0^+$. 

The evaluation for $A$=146~\cite{Data146}, in the section for the $^{146}$La nucleus, contains electron conversion coefficients for several transitions. Upon request from the authors of this work, the online version of NNDC now shows these conversion coefficients have been determined to come from theoretical HSICC calculations and are not experimental values. As such, they will not be used in this discussion.



\begin{figure*}
\includegraphics[width=0.99\textwidth, keepaspectratio]{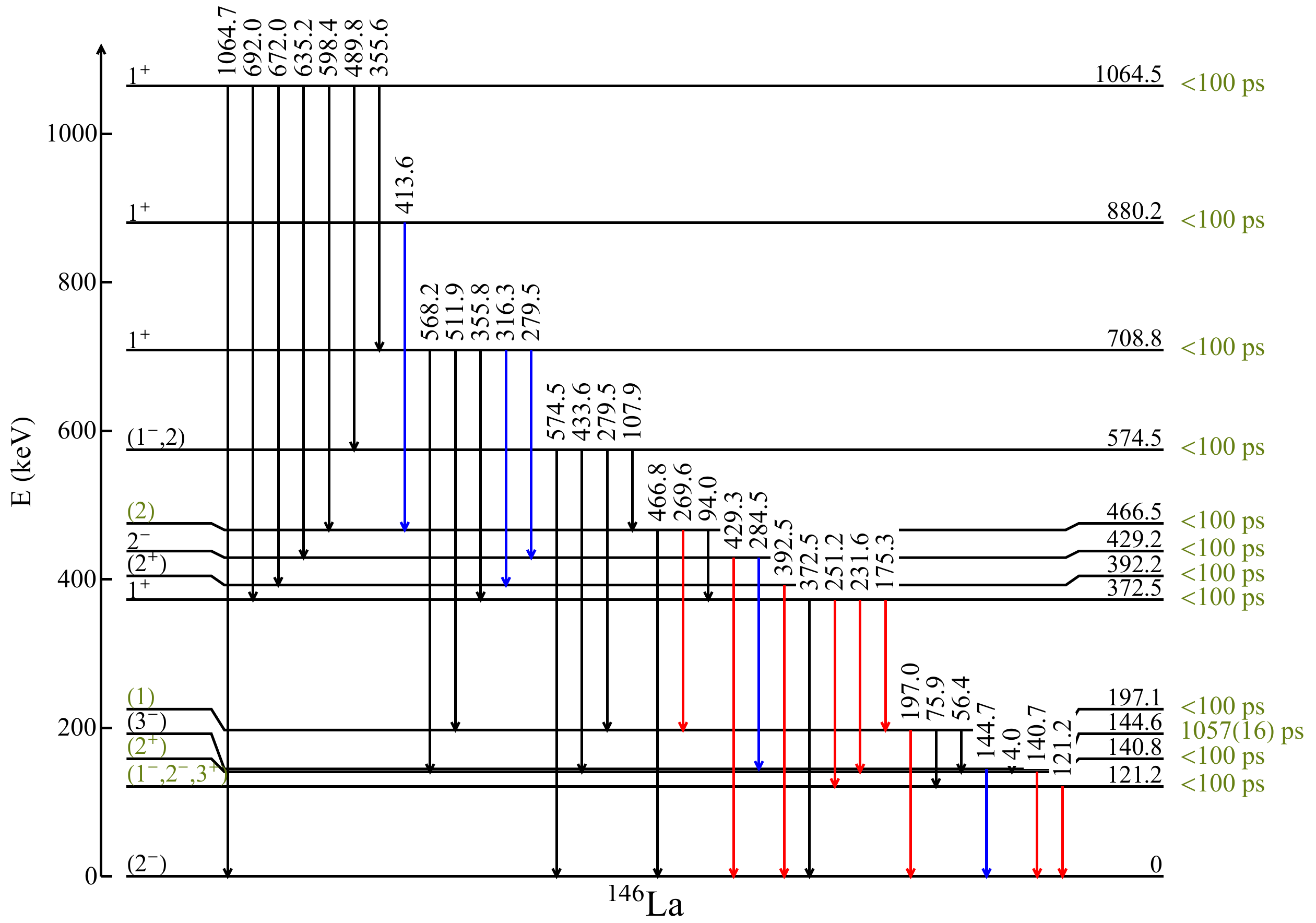}
  \caption{\textit{(Color online)} Same as Fig.~\ref{fig:145Ba-scheme} for the partial level scheme of $^{146}$La populated in the $\beta^-$ decay of $^{146}$Ba~\cite{Data146}.}
  \label{fig:146La-scheme}
\end{figure*}

A partial level scheme is presented in Fig.~\ref{fig:146La-scheme}.

\textbf{The 121.16-keV level \soutthick{$1^-,2^-$} $(1^-,2^-, 3^+)$:} Our results are compatible with the transition being either pure E1 or M1/E2 mixed character, although the M1 component would be significantly enhanced. Chung {\it et al.}~\cite{Chung1985} suggested that the results from angular correlations (372-121-0 keV states cascade), favoured the $1^+ \rightarrow (1^-,2^-) \rightarrow (2^-)$ cascade. However, given the relatively large uncertainties in the $A_{22}$ and $A_{44}$ parameters, spin 2 and 3 can also reproduce the data. The lack of direct $\beta$ feeding yields log$(ft)>7.9$, a strong indication that this is a forbidden transition. This discards the $0^+$ and $1^+$ possibilities. $I^\pi=3^-$ and above can be discarded by the strong transition from the $1^+$ 251.5-keV state. The measured lifetime also discards the possibility of the 121-keV transition having pure E2 character, thus ruling out this state as $I^\pi = 0^-$. The 75.9-keV transition from the 197-keV state does not help to rule out the $I^\pi = 3^+$ option due to the lack of parity assigment for the state (see details below). Therefore, $1^-, 2^-, 3^+$ are the most likely possibilities.


\textbf{The 140.85-keV level \soutthick{$(2^-_2)$} $(2^+_1)$:} The 140.7-keV transition is the only one observed to depopulate this state and, by employing angular correlations with the 232-keV transition, it was assigned $\delta=0.66(13)$~\cite{Chung1985}. The upper limit for the lifetime observed for this level yields an nonphysical B(E2)$>376$~W.u. for the E2 component of this transition. The mixing ratio needs to change to $\delta\sim0.05$ (nearly $4.5\sigma$ away) to obtain a more physical \textit{lower limit} of B(E2)$>50$~W.u., making the current interpretation of the angular correlation results highly unlikely. The only way to reconcile the results from this work and those of Ref.~\cite{Chung1985} is by changing the parity of this state to positive and thus have $I^\pi=2^+$. Angular correlations are insensitive to the parity of states, so only the mixing ratio of the involved transitions will be modified. If now the cascade is $1^+ \rightarrow 2^+ \rightarrow 2^-$, the first transition would have (M1+E2) character and the second one E1. A mixing ratio $\delta=0.39_{-5}^{+6}$ for the first would reproduce the $A_{22}$ and $A_{44}$ coefficients reported in Ref.~\cite{Chung1985}. $I^\pi=0, 1^+$ are discarded for this level by the lack of direct $\beta$-feeding and any $I>2$ would not fit the angular correlations~\cite{Chung1985}.

The angular correlation from Chung \textit{et al.}~\cite{Chung1985} for the 298-140-0 cascade is slightly harder to reconcile. The spin-parity of the 439-keV state is firmly (although it should be at least tentative) established as $1^-$ from log(\textit{ft})=6.1. However, the evaluators from Ref.~\cite{Data146} warn that, due to the lack of firm multipolarity assignments for the $\gamma$-transitions in levels below 1~MeV, the $I_\beta$ and the corresponding log(\textit{ft}) are not reliable. This, added to a potential \textit{Pandemonium} effect, leads to the interpretation of the direct $\beta$-feeding of the 439-keV state as a lower limit. In this case, only an allowed transition can be ruled out (and barely so). If $I^\pi =2^+$ is assumed for this state (second forbidden decay from the $0^+$ mother state), the cascade would be $2^+ \rightarrow 2^+ \rightarrow 2^-$, thus involving a (M1+E2) and an E1 transition. A small mixing ratio $\delta=0.09(8)$ would suffice to reproduce the observed angular correlation. The branching ratios of the transitions depopulating this state are difficult to interpret following either Chung and collaborators or our spin assignment. Considering all of these points a tentative assignment of $(2^+)$ is suggested for the 140.85-keV state. The alternative $I^\pi =(2^-)$ would make the measured angular correlation, T$_{1/2}$ and $I_\beta$ incompatibles.

In a single-particle model, it could be expected that the unpaired proton in $^{146}$La occupies the positive-parity $g_{7/2}, d_{5/2}$ orbitals, while the unpaired neutron occupies the negative-parity $f_{7/2}, p_{3/2}$ ones, thus yielding negative-parity states. To create a positive-parity state, the unpaired proton should be promoted to the $h_{11/2}$ orbital or the neutron to the $i_{13/2}$ one. If confirmed that this state has indeed positive parity, it would be a strong indication of significant deformation (at least $\epsilon \sim 0.15$) in this nucleus, needed for particles to occupy the latter orbitals at such low energies.


\textbf{The 144.6-keV level $(3^-)$:} The total intensity (I$_\gamma$+I$_e$) for the 4.0-keV transition is compiled in~\cite{Data146}, corrected from the data reported in~\cite{Chung1985}. Table~\ref{tab:transition-strength-146} shows that the pure B(E2) strength is almost a factor of 10 larger than the RUL~\cite{Endt1981}, hence discarding the presence of any significant E2 component. The RUL for E1 transitions is, strictly, B(E1)$10^{-2}$~W.u., but that corresponds to the tail of the distribution, making it very unlikely that this 4.0-keV transition indeed has an E1 character. Lastly, the RUL for M1 is B(M1)$<1$~W.u. Although the extracted B(M1)=$1.18(1)\cdot 10^{-1}$~W.u. is within the RUL, once again this corresponds to the end tail of the distribution with typical values B(M1)$\sim 10^{-2}$~W.u, making this hypothetical M1 a highly enhanced transition. Since the B(E2) and B(M1) values for the 144.7-keV transition seem reasonable, it appears unlikely that the measured lifetime (see Fig.~\ref{fig:lifetimes}) is the problem of the calculated 4.0-keV strength. Studies on the accuracy of electron conversion coefficients are limited to the \textit{K} and \textit{L} shells, since they are typically the only ones that can be observed~\cite{Kibedi2008}. For this 4.0-keV transition, only electrons in the \textit{M} shell and above are energetically possible, and consequently the quality of the theoretical calculations is not guaranteed. Nevertheless, any hypothetical deviation of $\alpha_M$ can hardly account for the factor of $\sim10$ needed to explain the extracted B(XL). It would be desirable, thus, to remeasure the intensity of this transition.
 

\textbf{The 197.03-keV level \soutthick{$(1^-)$} $(1)$:} The compilation~\cite{Data146} marks this state as a tentative $(1^-)$ character, based on a recalculated log(\textit{ft}) value of 5.65(11), significantly lower than the measured by Chung \textit{et al.}~\cite{Chung1985}. However, this log(\textit{ft}) value does not allow to distinguish between an allowed or first forbidden transition in this nucleus, indicating $I=0,1$. Therefore the assignment of the negative parity is uncalled for. Chung \textit{et al.}~\cite{Chung1985} also performed angular correlations, but the results shown, and the authors themselves so stated, that the measured A$_2$ coefficient does not impose restrictions on the possible spins of this level. The measured lifetime limit only helps to discard any significant E2 (or higher) component of the 56.4- and 197.0-keV transitions, imposing the limit $1 \leq I \leq 3$. Therefore, $I^\pi=1$ should be considered, without constrains on the possible parity.



\textbf{The 372.53-keV level $1^+$:} As mentioned before, the spin and parity of this state was established by the presence of direct $\beta$ feeding. The measured half-life upper limit yields average B(E1) lower limits for all transitions, but discard any possibility of M2 character. 

\textbf{The 392.6-keV level $(2^+)$:} This level is depopulated by the 392.5-keV transition to the $(2^-)$ g.s. and was proposed to have E1+M2 character. From angular correlations, a mixing ratio of $\delta=-0.28(10)$ was suggested~\cite{Chung1985}. This is in clear contradiction with the measured upper limit of T$_{1/2}<100$~ps that discards any significant M2 component. The lack of direct $\beta$ feeding to this level discards the assignment of an allowed transition. On the other hand, this level is populated by several higher-energy $1^+$ states, suggesting $I^\pi=(2,3^+)$. A more likely scenario is that the 392.5-keV transition has a pure E1 character and the measured mixing corresponds to the transitions from the feeding $1^+$ states (thus discarding the $3^+$ option). In the case of the 709-392-0 cascade, $\delta \sim -0.27$ for the M1+E2 316-keV transition would reproduce the measured angular distribution~\cite{Chung1985}.



\textbf{The 466.5-keV level \soutthick{$2^+$} $(2)$:} This state appears as firmly established in the evaluation~\cite{Data146}, citing the angular correlation work from Chung \textit{et al.}~\cite{Chung1985}. Specifically, it relies on the $\gamma$-ray cascade of the $880 \rightarrow 466 \rightarrow 197$~keV states transitions, that is proposed to be a $1^+ \rightarrow 2^+ \rightarrow (1^-)$ sequence. However, in that work the authors did not assign any spin-parity to the state, as their results were inconclusive. 

A log(\textit{ft}) value $>6.4$ suggests that the $\beta$ decay is either a first or second forbidden transition. A second forbidden decay would mean $I^\pi=2^+, 3^+$, whereas if the decay is a first forbidden it would indicate $I^\pi=1^-, 2^-$.

The lifetime upper limit yields B(E2)$>290$~W.u. for both the 94.0-keV transition to the $1^+_1$ state and the 139.8-keV transition to the (3) state. This lower limit, so close to the RUL, strongly suggests that this state cannot have $I=1$ or 3. Therefore, it is proposed $I=(2)$, but without constraints on the parity.

\textbf{The 708.8-keV level $1^+_2$:} The spin-parity was firmly assigned by log($ft$)=5.07(4) in $\beta$ decay~\cite{Chung1985}. Although only an upper limit was measured for the lifetime, it is noteworthy how the B(E1) strength of the 279.5-keV $1^+_2 \rightarrow 2^-$ is enhanced by two orders of magnitude with respect to the 511.9- and 568.2-keV transitions (or alternatively, these two transitions are suppressed). 

\section{\label{sec:conclusion}Conclusion}


This paper reports on measurements of excited-state lifetimes for $^{144, 145, 146}$Ba and $^{145,146}$La populated in the $\beta^-$ and $\beta^-$-$n$ decay of $^{145,146}$Cs. Seven of these lifetimes are measured for the first time and there is very good agreement with previous values in the other cases. The experiment was performed at TRIUMF-ISAC using the GRIFFIN spectrometer and employing electronic fast-timing techniques.

The measured lifetimes were used to calculate B(XL) values and from them extract the multipolarity character of the transitions, with limited success. In many cases, the new lifetimes measured in this paper contradict the spin-parity assignments of previous works, including some that were established as firm. New tentative $I^\pi$ are suggested.

In the case of $^{145}$Ba, a candidate for the parity doublet of the ground state is tentatively proposed for the first time. If confirmed, it will show that the drop in $D_0$ observed for $^{146}$Ba, is also present in this odd-isotope. A greatly enhanced B(E2) value was also observed for the transition connecting the first excited and ground states of this nucleus. The origin of this large B(E2) remains to be explained.

New information for $^{146}$La provides a strong indication of a low-lying positive-parity state at only 140.85~keV. If confirmed, this would point to a sudden onset of deformation in the neutron-rich La.

Despite the wealth of information available for these isotopes, few of their states have firmly established spin-parity. This greatly hinders reaching any firm conclusion on their nuclear structure or meaningful comparison with theoretical calculations. Thus, despite the predictions, the permanent octupole deformation in the region remains an incognita for these odd and odd-odd nuclei. As an example to illustrate this problem, a recent work in which $^{143}$Ba was measured using Coulomb excitation was unable to draw firm conclusions on static octupole deformation despite having measured B(E3) values~\cite{Morse2020}. The authors argued that several low-lying states had unknown or tentative spin and parities, making any interpretation highly speculative.

This work highlights the need for future experiments with high-precision angular correlations and conversion electron spectroscopy in order to elucidate the low-lying structure of these isotopes. This is especially important before Coulomb excitation measurements can be attempted.

\textit{Note added in Proof.} Recently, the authors were made aware of a recent publication~\cite{Cardona2021} reporting the measurement of several lifetimes in $^{145}$La. There is perfect agreement between the results of both experiments, which is a strong confirmation of the present results.

\begin{acknowledgments}
The authors would like to thank B. Singh, from MacMaster University, and T. Kibedi, from Australian National University, for insightful discussion of the results. The GRIFFIN infrastructure has been funded jointly by the Canada Foundation for Innovation, TRIUMF and the University of Guelph. TRIUMF receives federal funding via a contribution agreement through the National Research Council Canada (NRC). C.E.S. acknowledges support from the Canada Research Chairs program. This work was supported in part by the Natural Sciences and Engineering Research Council of Canada (NSERC).
\end{acknowledgments}

\bibliographystyle{apsrev}
\bibliography{Ba_bibliography}

\end{document}